\documentclass[showpacs,aps,twocolumn,nofootinbib,amsmath,superscriptaddress,secnumarabic]{revtex4-1}
\usepackage{epsfig,multirow,pstricks,braket,hyperref}
\hypersetup{colorlinks=true,linkcolor=blue,citecolor=magenta,filecolor=magenta,urlcolor=cyan}

\def\({\left(}
\def\){\right)}
\def\bal#1\eal  {\begin{align} #1 \end{align}}
\def\bsp#1\esp{\begin{split}#1\end{split}}
\newcommand{\be} {\begin{equation}}
\newcommand{\ee} {\end{equation}}

\newcommand{\ud} {\mathrm{d}}
\newcommand{\pd} {\partial}
\newcommand{\mc} {\mathcal}
\newcommand{\bfr} {{\bf r}}
\newcommand{\ai}{{\alpha}}
\begin{document}
\title{ Positivity in electron-positron scattering: testing the axiomatic quantum field
  theory principles and probing the existence of UV states}
\author{Benjamin Fuks}
\email{fuks@lpthe.jussieu.fr}
\affiliation{Sorbonne Universit\'e, CNRS, Laboratoire de Physique Th\'eorique et
  Hautes \'Energies, LPTHE, F-75005 Paris, France}
\affiliation{Institut Universitaire de France, 103 boulevard Saint-Michel,
  75005 Paris, France}
\author{Yiming Liu}
\email{liuym@ihep.ac.cn}
\affiliation{Institute of High Energy Physics, Chinese Academy of Sciences,
Beijing 100049, China}
\author{Cen Zhang}
\email{cenzhang@ihep.ac.cn}
\affiliation{Institute of High Energy Physics, Chinese Academy of Sciences,
Beijing 100049, China}
\affiliation{
School of Physical Sciences, University of Chinese Academy of Sciences, Beijing
100049, China
}
\affiliation{Center for High Energy Physics, Peking University, Beijing 100871, China}
\author{Shuang-Yong Zhou}
\email{zhoushy@ustc.edu.cn}
\affiliation{Interdisciplinary Center for Theoretical Study, University of Science and Technology of China, Hefei, Anhui 230026, China}
\affiliation{Peng Huanwu Center for Fundamental Theory, Hefei, Anhui 230026, China}

\preprint{USTC-ICTS/PCFT-20-25}

\begin{abstract}
  We consider the positivity bounds on dimension-8 four-electron operators and
  study two related phenomenological aspects at future lepton colliders. First,
  if positivity is violated, probing such violations will
  revolutionize our understanding of the fundamental pillars of quantum field
  theory and the $S$-matrix theory. We observe that positivity violation at scales
  of 1--10~TeV can potentially be probed at future lepton colliders even if
  one assumes that dimension-6 operators are also present. Second, the positive
  nature of the dimension-8 parameter space often allows us to either directly
  infer the existence of UV-scale particles together with their quantum
  numbers or exclude them up to certain scales in a
  model-independent way.  In particular, dimension-8 positivity plays an
  important role in the test of the Standard Model. If no
  deviations from the Standard Model are observed, it allows for simultaneous exclusion
  limits on all kinds of potential UV-complete models. Unlike the
  dimension-6 case, these limits apply regardless of the UV model setup and
  cannot be removed by possible cancellations among various UV contributions.
  This thus consists of a novel and universal test to confirm the Standard Model.
  We demonstrate with realistic examples how all the previously mentioned
  possibilities, including the test of positivity violation, can be achieved.
  Hence, we provide an important motivation for studying dimension-8 operators
  more comprehensively.
\end{abstract} 

\maketitle
%\tableofcontents

\section{Introduction}

After assuming that the UV completion of the Standard Model Effective Field
Theory (SMEFT) satisfies $S$-matrix and quantum field theory (QFT) axiomatic
properties, such as Lorentz invariance, unitarity, analyticity, and locality,
one can show that the SMEFT dimension-8 Wilson coefficients must satisfy the
so-called positivity bounds~\cite{Zhang:2018shp, Bi:2019phv, Zhang:2020jyn,
Bellazzini:2018paj, Remmen:2019cyz, Remmen:2020vts} (see, {\it e.g.},
Refs.~\cite{Adams:2006sv, deRham:2017avq, deRham:2017zjm, Arkani-Hamed:2020blm, Bellazzini:2016xrt} and
the references therein for generic discussions). While these
bounds could guide experimental searches for physics beyond the Standard Model
(SM), one might, conversely, consider them to test the
axiomatic principles of QFT experimentally~\cite{Distler:2006if}. If the measurements of the values
of the Wilson coefficients violate the positivity bounds,
the underlying UV model must violate at least one of those principles. Therefore,
new ideas beyond conventional model building approaches
are required.

Another interesting feature of the dimension-8 coefficient space is that, by
exploiting its positive nature, we can either infer the existence of UV states
and their quantum numbers~\cite{Zhang:2020jyn} or exclude them. This originates from
the positivity bounds that carve out a geometric object in the parameter space
of the Wilson coefficients, namely a convex cone whose ``edges'' (to be more
precisely defined later) are closely related to the properties of the UV states
lying in specific irreducible representations (irreps) of the SM symmetry group.
This relation, when combined with the positive nature of the dimension-8
coefficient space, can often provide striking information about the new physics
states living in the UV.

As an extreme example, one could imagine that one
succeeds in measuring the vector of dimension-8 Wilson coefficients with
sufficient precision. If this vector coincides with one of the ``edges'' of the
cone, the UV states generating the corresponding operators can be uniquely
determined in the sense that they {\it must} all lie in a single irrep of the SM
symmetries. If the UV completion is assumed to be tree-level and weakly coupled,
one concludes that the underlying theory must be a ``one-particle
extension'' of the SM. This provides an answer to the ``inverse
problem''~\cite{ArkaniHamed:2005px,Dawson:2020oco, Gu:2020thj} that can be
summarized as follows. Given the measured values of the coefficients at the
electroweak scale, how can we possibly determine the nature of the new physics
beyond the SM? Similarly, if the measurement agrees with the SM value to
a sufficient precision, one can simultaneously exclude the existence of any
potential new physics state up to certain scales.  This
exclusion is guaranteed by the positive nature of the Wilson coefficients and
cannot be removed by arranging the UV states in specific patterns that cancel
each other's effects. Therefore, this would provide a model-independent
confirmation of the SM, which is not possible when truncating the SMEFT at
dimension-6.

Dimension-8 SMEFT operators~\cite{Li:2020gnx,Murphy:2020rsh} have recently
attracted increasing attention, in particular as the LHC accumulates more
data. Various motivations for going beyond a truncation of the SMEFT Lagrangian
at the dimension-6 level have correspondingly been presented, for example, in
Refs.~\cite{Liu:2016idz,Azatov:2016sqh,Bellazzini:2017bkb,Ellis:2018cos,
Bellazzini:2018paj,Hays:2018zze,Ellis:2019zex,Remmen:2020vts,Ellis:2020ljj}.
In addition, observables that can be used to disentangle the effects of dimension-8 operators
from those of dimension-6 have been proposed and studied, as for example, in
Refs.~\cite{Alioli:2020kez,Gupta}. However, positivity-related topics, including
the possible tests of its violation and the option of inferring/excluding the
existence of states in the UV, have not been discussed using realistic
phenomenological examples.

This study aims to present some initial results in this direction in
the context of future lepton colliders. In particular, we are interested in the
following questions:
\begin{enumerate}
  \item To what extent can we test any potential violation of the positivity
    bounds?
  \item For realistic measurements (including the associated experimental
    errors), to what extent can we learn about the existence of UV states and
    their properties using the positive nature of the dimension-8 coefficient
    space?
\end{enumerate}
The first point has been discussed in Ref.~\cite{Distler:2006if} but not in the
SMEFT framework, and no realistic collider analysis has been presented.
In contrast, the second point has not been discussed in literature.

Several proposals for a future electron-positron machine are currently
discussed, including the CEPC~\cite{CEPCStudyGroup:2018ghi}, FCC-ee~\cite{
Abada:2019lih,Abada:2019zxq}, ILC~\cite{Bambade:2019fyw,Asner:2013psa}, and
CLIC~\cite{deBlas:2018mhx} projects. These colliders present an ideal means to
perform high accuracy measurements, particularly as they are planned to be operated at
various center-of-mass energies. Thus, they could allow us to
distinguish the effects of dimension-8 operators from those of dimension-6 on a large set of
observables, thus creating new opportunities to access to information on the
SMEFT dimension-8 operators.\footnote{In addition to the energy dependence, the
angular momentum can also be used as a discriminant~\cite{Alioli:2020kez}.}

For a first step in this direction, we consider the simplest $2\to 2$
process that could occur at a lepton collider and that is expected to be
one of the most accurately probed processes, $e^+e^-\to e^+e^-$, and investigate
the impact of the SMEFT four-fermion operators. We ignore other $e^+e^-\to ff$
channels with other final-state fermion species, as the corresponding positivity
bounds involve not only $e^2f^2$ operators but also $e^4$ and $f^4$ operators.
On the contrary, $e^+e^-\to e^+e^-$ is self-contained in the sense that only
$e^4$ operators are relevant, and their positivity bounds do not involve other
operators at leading order. A more comprehensive study including more
operators and processes can be undertaken, but we leave it for the future.

The remainder of this paper is organized as follows. In Section~\ref{sec:ops}, we list the $e^4$
dimension-6 and dimension-8 operators relevant to our study. In
Section~\ref{sec:positivity}, we derive the positivity bounds on these operators
by using the elastic scattering of arbitrarily superposed states. We propose a
variable to quantify the amount of potential positivity violation in
Section~\ref{sec:violation} to connect the collider reaches with the
underlying physics and discuss its possible interpretations in
Section~\ref{sec:interpretation}. Then, in Section~\ref{sec:inverse}, we briefly
discuss how to infer/exclude the existence of UV states using dimension-8
positivity. Subsequently, in Section~\ref{sec:collider}, we study the phenomenological
aspects of positivity for several $e^+e^-$ collider scenarios. Finally, we summarize our
main findings in Section~\ref{sec:summary}.

\section{Effective operators}
\label{sec:ops}
The SMEFT Lagrangian is generically defined as
\begin{flalign}
 \mathcal{L}_\mathrm{SMEFT}=
 \mathcal{L}_\mathrm{SM}+\sum_i\frac{C_i^{(6)}}{\Lambda^2}O_i
 +\sum_i\frac{C_i^{(8)}}{\Lambda^4}O_i+\dots
\end{flalign}
where the $C_i$ parameters represent the various Wilson coefficients associated
with the higher-dimensional operators $O_i$, and $\Lambda$ denotes the
cutoff scale of the theory. At the tree level, two classes of effective
operators are relevant for $e^+e^- \to e^+e^-$ scattering. The first one
involves four electron fields
(four-fermion operators), and the second one involves two electron fields ({\it e.g.},
the operators affecting the $\bar eeZ$ or $\bar ee\gamma$ vertices) or fewer
({\it e.g.}, the operators modifying the electroweak boson two-point functions).
For a feasibility study, we solely focus in this work on four-fermion operators,
assuming that the other potentially relevant operators can be determined or
constrained by the study of other $e^+e^-\to X$ channels. We also
ignore any possible loop-level correction within the SMEFT framework.

There are three four-fermion $e^4$ operators that arise at
dimension-6~\cite{Grzadkowski:2010es},
\be\bsp
 &O_{ee}=( \bar e \gamma^\mu e )\ ( \bar e \gamma_\mu e)\ ,\\
 &O_{el}=( \bar e \gamma^\mu e )\ ( \bar l \gamma_\mu l)\ ,\\
 &O_{ll}=( \bar l \gamma^\mu l )\ ( \bar l \gamma_\mu l)\ ,
\esp\label{eq:dim6}\ee
and all of which provide independent contributions to $e^+e^-\to e^+e^-$ scattering.
We use the notations $C_{ee}$, $C_{el}$ and $C_{ll}$ below to
denote their dimensionless Wilson coefficients.

The full basis of dimension-8 operators has been presented recently~\cite{
Li:2020gnx,Murphy:2020rsh}. Three types of four-electron operators are relevant to our
study and are of the forms $\Psi^4D^2$ (four-fermion operators including two
derivatives), $\Psi^4H^2$ (four-fermion operators including two extra Higgs
fields), and $\Psi^4DH$ (four-fermion operators including one derivative and one
extra Higgs field).

In this study, we are mainly interested in operators $\Psi^4D^2$ of the
first category, as they are subject to positivity bounds. There are five such
independent operators for which we choose the following basis:
\be\bsp
 &O_{1}=\partial^\alpha(\bar e \gamma^\mu e) \partial_\alpha (\bar e \gamma_\mu e)\ ,\\
 &O_{2}=\partial^\alpha(\bar e \gamma^\mu e) \partial_\alpha (\bar l \gamma_\mu l)\ ,\\
 &O_{3}=D^\alpha(\bar e l)\ D_\alpha(\bar l  e), \\
 &O_{4}=\partial^\alpha(\bar l \gamma^\mu l)\ \partial_\alpha(\bar l \gamma_\mu l)\ ,\\
 &O_{5}=D^\alpha(\bar l \gamma^\mu \tau^I l)\ D_\alpha(\bar l \gamma_\mu \tau^I l)\ ,
\esp\ee
where the $\tau^I$ matrices are the Pauli matrices. The operators $O_{4}$ and $O_{5}$
contribute identically to $e^+e^-\to e^+e^-$; therefore, this process is
only sensitive to the four independent coefficient combinations $C_1$, $C_2$, $C_3$,
and $C_4+C_5$, where $C_i$ denotes the Wilson coefficient associated with the
operator $O_i$. Consequently, in the collider discussions below, we will
always set $C_5=0$. This is equivalent to restricting our discussion to
four independent degrees of freedom in the considered process,
ignoring the fact that the left-handed electron and neutrino live in the same
$SU(2)_L$ doublet.

Similar to a model-independent SMEFT framework, the other two classes of
dimension-8 operators should be included as well. However, in the context of
$e^+e^-\to e^+e^-$ scattering, the $\Psi^4H^2$ operators act like dimension-6
operators (after replacing the two Higgs fields by their vacuum expectation values).
Thus, these operators can only be disentangled from the dimension-6 ones when more
observables are included, such as those related to neutrino DIS experiments. In
our study, these operators can be fully captured by shifting the three dimension-6
coefficients of Eq.~\eqref{eq:dim6}. Once the latter are marginalized over, they
have no impact on the determination of the dimension-8 operators of the
first category $\Psi^4D^2$. Finally, the $\Psi^4DH$ operators can be omitted by
assuming a $U(3)^5$ flavor symmetry.

In summary, our collider analysis only incorporates the effects of the
dimension-8 operators of the first type $\Psi^4D^2$. In the following, we
frequently refer to a vector notation for the Wilson coefficients,
\be
 \vec C^{(6)}=(C_{ee},C_{el},C_{ll}), \qquad
 \vec C^{(8)}=(C_1,C_2,C_3,C_4)\ .
\label{eq:Wilvec}\ee
This allows for a parameterization of (differential) cross sections
up to $\mathcal{O}(\Lambda^{-4})$ as,
\be
  \sigma=\sigma_\mathrm{SM} +\!
  \sum_i\frac{C_i^{(6)}}{\Lambda^2}\sigma^{(6)}_i
  \!+\!\sum_i\frac{\big[C_i^{(6)}\big]^2}{\Lambda^4}\sigma^{(6)}_{ii}
  \!+\!\sum_i\frac{C_i^{(8)}}{\Lambda^4}\sigma^{(8)}_i \ ,
\label{eq:poly}\ee
where $C_i^{(6)}$ and $C_i^{(8)}$ run through all the components of the
vectors $\vec C^{(6)}$ and $\vec C^{(8)}$. This expression includes the fact
that, in the (adopted) limit of $m_e\to0$, there is no interference between two
different dimension-6 operators. We have computed the different $\sigma$ terms
both analytically and by using {\sc FeynRules}~\cite{Alloul:2013bka} to
generate a UFO library~\cite{Degrande:2011ua} to be used within
{\sc MadGraph5\_aMC@NLO}~\cite{Alwall:2014hca}. To assess the impact of
truncating the $1/\Lambda$ expansion, we also include the next order terms in
some of our results. Equivalently, we add interference terms involving a
$C^{(6)}_i C^{(8)}_j$ product and the quadratic contributions in the
$C^{(8)}_i$ coefficients.

\section{Positivity bounds}
\label{sec:positivity}

Positivity bounds can be derived from a dispersion relation and the optical
theorem, which are based on fundamental QFT principles including unitarity,
analyticity, locality, and Lorentz invariance. This is a very active field with
a vast and growing literature, and in this regard, we refer to Refs.~\cite{Adams:2006sv,
deRham:2017avq, deRham:2017zjm, Arkani-Hamed:2020blm} and the references therein and to
Refs.~\cite{Zhang:2018shp, Bi:2019phv, Zhang:2020jyn, Bellazzini:2018paj,
Remmen:2019cyz, Remmen:2020vts} for specific SMEFT applications.

The conventional approach to derive positivity bounds utilizes forward and
elastic scattering amplitudes (see, {\it e.g.}, Ref.~\cite{Zhang:2018shp}).
Briefly, it requires that the second order $s$-derivative of the elastic
amplitudes (with poles subtracted) be positive. For instance, for the
process considered in this study, it is given by
\begin{flalign}
\label{disprelation}
	\frac{\ud^2}{\ud s^2}M(e^+e^-\to e^+e^-)\ge0  .
\end{flalign}
The l.h.s.~of this equation consists of a linear combination of the considered
$\Psi^4D^2$ operator coefficients, and the above requirement determines the sign
of this combination. More specifically, different bounds can be derived by
choosing different fermion species and chiralities, as for example with
\begin{itemize}
\item $M(e_R e_R\to e_R e_R)$: $C_1\le0$;
\item $M(e_L e_L\to e_L e_L)$: $C_4+C_5\le0$;
\item $M(e_R \bar e_L\to e_R \bar e_L)$: $C_3\ge0$;
\item $M(e_L \nu_L\to e_L \nu_L)$: $C_5\le0$.
\end{itemize}
However, the above list of bounds is not complete, as, by defining states through
the superposition of different flavors and chiralities, one can consider extra
elastic scattering processes~\cite{Wang:2020jxr}. According to this approach, the best
bounds are derived in Appendix~\ref{sec:a1} from the scattering amplitudes
presented in Appendix \ref{sec:app2},
\begin{flalign}
	&C_1\le0  \label{eq:bound0},\\
	&C_4+C_5\le0,\\
	&C_5\le0,\\
        &C_3\ge0,\\
	&2\sqrt{C_1(C_4+C_5)}\ge C_2, \label{eq:bound1}\\
	&2\sqrt{C_1(C_4+C_5)}\ge-(C_2+C_3). \label{eq:bound2}
\end{flalign}
While the first four bounds are obtained without considering any superposition,
those of Eqs.~\eqref{eq:bound1} and \eqref{eq:bound2} arise from the
superpositions
\be\bsp
	| f_\pm \rangle \equiv &\frac{(C_4+C_5)^{1/4}}
	{\left[(C_4+C_5)^{1/2}+C_1^{1/2}\right]^{1/2}} | e_R\rangle 
	\\&\quad  \pm
	\frac{C_1^{1/4}}
	{\left[(C_4+C_5)^{1/2}+C_1^{1/2}\right]^{1/2}}  |\bar e_L\rangle  ,
\esp\ee
and the scattering processes $f_\pm f_\mp \to f_\pm f_\mp$ and $f_\pm f_\pm \to
f_\pm f_\pm$, respectively. Moreover, these two bounds are homogenous and written
as quadratic inequalities of the Wilson coefficients.

The above approach is sufficient for the study of the four-fermion operators
considered in this paper. However, in general, it is insufficient to obtain
the best possible bounds. Accordingly, a new and better approach has
been proposed recently~\cite{Zhang:2020jyn}. The idea is to construct the allowed
Wilson coefficient parameter space region directly as a convex cone, which is a convex
hull of its extremal rays, and the latter can be identified using group
theoretical considerations.

This new approach has at
least two advantages. First, it always provides the tightest constraints available
from the dispersion relation. They may be tighter than those that can be obtained by
relying on the conventional elastic positivity approach, as for example, for the
scattering of a pair of $W$-bosons. Second, more relevant to this study (see
Section~\ref{sec:inverse}), it reveals a connection between the positivity
bounds and the existence of new physics states in the UV. In order for this
connection to be manifest, one needs to determine the exact shape of the
parameter space allowed by the bounds, which cannot always be achieved with the
conventional approach for complicated cases. We have verified that,
for the operators under consideration, the two methods yield the same set of
bounds of Eqs.~\eqref{eq:bound0}--\eqref{eq:bound2}.

We devote the remainder of this section to a discussion on the possible violation of
positivity and its physical implications.

\subsection{Quantifying positivity violation}
\label{sec:violation}

A positivity violation would imply a breakdown of the fundamental principles of
QFT. Hence, if such a violation is observed at a future collider, it would be
mandatory to study the physics behind it. To this end, we first need a
model-independent way to quantify the observed amount of violation, which can be
connected later to possible physics scenarios.

The physical quantity to consider for probing a potential positivity violation
is the second order $s$-derivative of the studied amplitude with poles
subtracted, $M(s, t=0)$. Introducing the explicit dependence on the fermion
mixing parameters $\epsilon_i=(a_i,b_i,c_i)$ of Eq.~\eqref{eq:mixing}, we can
define
\begin{flalign} \label{DeltaDef}
 -\Delta^{-4}\equiv
 \min\left[ \min_{\epsilon_1,\epsilon_2} \frac{1}{2}
    \frac{\ud^2M(s,t=0)(\epsilon_1,\epsilon_2)}{\ud s^2},0\right]\ ,
\end{flalign}
so that $\Delta$ has a mass dimension 1 and indicates the maximal amount of
positivity violation reached when the $\epsilon_i$ mixing parameters vary. Thus,
if positivity is always satisfied, $\Delta=\infty$. For the
amplitude to be physical, we impose the constraints $|\epsilon_{1,2}|=1$.
Generically, there will be several contributing operators, so that it is
convenient to consider the level of the amplitude directly instead of that of
the Wilson coefficients. Ultimately, the amplitude contains the essential
physical information of the theory, whereas the operators in the Lagrangian are
subject to ambiguities originating from field redefinitions.

While the physical interpretation of the $\Delta$ quantity will be discussed
in Section~\ref{sec:interpretation}, we briefly comment below on its relation
with the scale of new physics $\Lambda_\mathrm{BSM}$. Let us assume that a set
of Wilson coefficients is measured, and both $\Lambda_\mathrm{BSM}$ and
$\Delta$ can be estimated/computed from these coefficients. Intuitively, we
expect the scale $\Delta$ to be larger, as a deviation from the SM does not
necessarily imply positivity violation. On the contrary, the latter implies that
beyond the SM (BSM) physics must exist.

One often estimates $\Lambda_\mathrm{BSM} = \Lambda/\sqrt[4]{C}$, where $C$
represents a typical dimension-8 Wilson coefficient. However, if multiple
coefficients are nonzero, it is instead more natural to infer
$\Lambda_\mathrm{BSM}$ from the amplitude that is physical and
basis-independent. It is desirable to use the largest possible amplitude
obtained when varying the initial-state and final-state superpositions,
\begin{flalign}
  \Lambda_\mathrm{BSM}^{-4}=\max_{\epsilon_1,\epsilon_2}\left|
  \frac{1}{2}\frac{\ud^2M(s,t=0)(\epsilon_1,\epsilon_2)}{\ud s^2}\right|\ ,
\end{flalign}
where the second-order derivative allows for the extraction of the dimension-8
deviations from the SM.
Hence, $\Lambda_\mathrm{BSM}\lesssim\Delta$ is as intuitively expected.

To compute the positivity violation measure $\Delta$, a simple option is to find
the set of $\epsilon_i$ mixing parameters that saturates to the
bounds of Eqs.~\eqref{eq:bound0}--\eqref{eq:bound2}. These bounds can be
considered as the supporting planes of a convex cone representing the set of
points in the Wilson coefficient parameter space that can be
UV-completed~\cite{Zhang:2020jyn}. Thus, for any point located outside
the cone, $\frac{1}{2}\frac{\ud^2M}{\ud s^2}$ is related to the distance to any
supporting plane of the cone (or in other words, the $\Delta$ parameter). This
yields
\be
\Delta^{-4}=\frac{\delta(\vec C^{8})}{\Lambda^4}\  ,
\label{eq:def2}\ee
with
\be\bsp
  &\delta(\vec C^{(8)})\equiv -\min\bigg[0,
    -4C_1,-4(C_4\!+\!C_5),C_3,-8C_5, \\ 
  &\quad
    \Theta[C_2 \!-\!2\max(C_1,C_4\!+\!C_5)]
     \frac{C_2^2\!-\!4C_1(C_4\!+\!C_5)}{C_1\!-\!C_2\!+\!C_4\!+\!C_5},\\
  &\quad
    \Theta[-C_2-C_3-2\max(C_1,C_4+C_5)]\\
  &\qquad
    \times\frac{(C_2+C_3)^2-4C_1(C_4+C_5)}{C_1+C_2+C_3+C_4+C_5}\bigg] ,
\esp\label{eq:def3}\ee
where $\Theta[x]$ is the standard Heaviside function. One can then use either
$\Delta^{-1}$ or $\delta(\vec C^{(8)})$ to assess the amount of positivity
violation. The former is
dimensionful and directly connected to the scale at which the fundamental QFT
principles are violated, whereas the latter is dimensionless and intended to be
combined with $\Lambda^{-4}$.

However, it is not necessarily sufficient to derive the $\epsilon_i$ values
yielding the six bounds of Eqs.~\eqref{eq:bound0}--\eqref{eq:bound2}.
{\it A priori}, there might exist
a supporting plane that provides a larger amount of
positivity violation, although as a positivity bound it can be positively
decomposed into a combination of Eqs.~\eqref{eq:bound0}--\eqref{eq:bound2}.
Therefore, it is redundant and discarded.
The rigorous way to estimate $\Delta$ is to rely on the
definition of Eq.~\eqref{DeltaDef} and minimize its right-hand side. We have
numerically checked that, for more than 90\% of the parameter space, the
difference between the exact method and the approximation of
Eq.~\eqref{eq:def3} is less than 10\%. Therefore, we consider the latter as a
convenient estimate.

\begin{figure}
  \begin{center}
    \includegraphics[width=\linewidth]{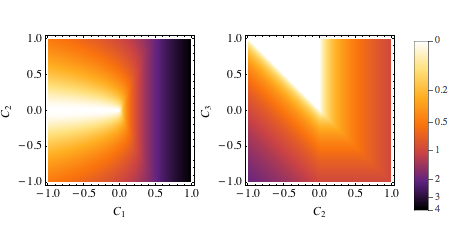}\\
    \includegraphics[width=\linewidth]{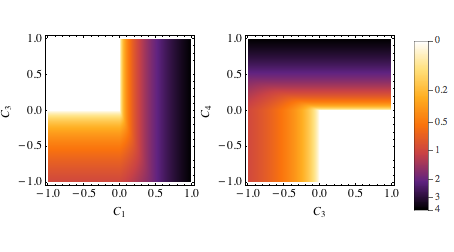}
    \caption{Amount of positivity violation in the studied dimension-8 Wilson
      coefficient parameter space. We consider several two-dimensional slices
      of the parameter space that we define by setting all Wilson coefficients
      but two to 0. Positivity violation is
      estimated through the quantity $\delta(\vec C^{(8)})$ of
      Eq.~\eqref{eq:def3}, so that the white areas correspond to regions
      compliant with positivity. \label{fig:violation}}
  \end{center}
\end{figure}

As an illustration, Figure~\ref{fig:violation} shows the dependence of
the quantity $\delta(\vec C^{(8)})$ on the different considered Wilson
coefficients. We focus on several two-dimensional slices of the parameter space
that we define by setting the irrelevant coefficients $C_i$ to 0. Thus,
the white areas for which $\delta(\vec C^{(8)}) = 0$ satisfy positivity. We can
observe that the amount of violation increases as one moves further away from the
positive regime.

\subsection{Physical interpretations}
\label{sec:interpretation}

What do possible violations of positivity bounds mean? It is known that the forward
positivity bounds are derived by assuming that the scattering amplitudes
computed in the UV-completed theory are unitary, Lorentz invariant, polynomially
bounded in momenta and analytical in the complex $s$ plane, apart from certain
poles and branch cuts. The unitarity of the $S$-matrix indicates that the
quantum mechanical probabilities of all possible scatterings add up to 1, which
results in the optical theorem, or that the imaginary part of the UV
amplitudes is positive in the physical region.

Violations of unitarity, such as the existence
of (bad) ghosts~\cite{Woodard:2015zca} in the UV, will lead to catastrophic
instabilities in the theory (and thus should be avoided), unless the ghosts only
appear in the effective field theory context and with a mass at or greater than
the cutoff scale. Lorentz invariance has been tested at very high energy scales
via various experiments, although certain properties are only weakly
constrained~\cite{Kostelecky:2008ts}. Analyticity is implied by causality and
can be proven to be valid at any perturbative order, although it has never been
proven non-perturbatively. The polynomial boundedness of the amplitudes
in the momentum space originates from locality, the lack of which would result in
ill-defined Fourier transforms and non-locality in real space.
In addition, polynomial boundedness, analyticity, and unitarity can be used to
prove the Froissart bound that implies that forward UV amplitudes should grow
slower than $s\ln^2 s$ when $s\to \infty$~\cite{Froissart:1961ux}. This allows
for the derivation of the dispersion relation (another important element to
derive the positivity bounds).

In terms of the Wilson coefficient parameter space, the observation of small
violations of the positivity bounds would indicate regions not too far away
from the positive regime ({\it e.g.}, the yellow/orange areas in
Figure~\ref{fig:violation}) and would imply that some of the fundamental
principles of QFT are violated at certain energy scales. A high experimental
precision is required to detect those small violations. In contrast, the
observation of a stronger violation of the positivity bounds would indicate
regions further away from the positive regime ({\it e.g.}, the purple/black
areas in Figure~\ref{fig:violation}) and imply violations of the
fundamental principles of QFT at more accessible energies.

On dimensional grounds, the quantity $\Delta$ introduced in
Section~\ref{sec:violation} regulates the amount of positivity violation. However,
how well does $\Delta$ connect to the actual positivity violation scale? To
investigate this issue, we consider a scenario in which the fundamental principles
of QFT are violated at some energy scale $\Lambda_*$. In this case, we can only
push the dispersion relation up to the scale $\Lambda_*$ in the complex $s$
plane, rather than to infinity where the semi-circular contour integrations
vanish by virtue of the Froissart bound (see, {\it e.g.},
Ref.~\cite{Bi:2019phv}). Such an earlier cutoff could be due, for instance, to
some new singularities at $|s|=\Lambda_*$ lying away from the real axis. If we
additionally assume that $\Lambda_*$ is parametrically greater than the mass
scale $M_{\rm BSM}$ of new particles living in the UV, the dispersion
relation can be written as
\be\bsp
  \frac12 \frac{\ud^2 M(s=2m_e^2,t=0)}{\ud s^2}  &=
    \int^{\Lambda_*^2}_{4m_e^2} \frac{\ud s'}{2\pi} \frac{\Im[M(s',0)]}{(s'-2m_e^2)^3}
    \\&~~~~
  + \int_{{\cal C}'}\frac{\ud s'}{2\pi i} \frac{M(s',0)}{(s'-2m_e^2)^3} ,
\esp\label{eq:disp}\ee
where ${\cal C}'$ denotes the two semi-circular contours at $|s|=\Lambda_*$. The
integral around the branch cuts along the real axis is positive by the usual
argument of positivity bounds, so that the violation of positivity could be
estimated by evaluating the integral along the ${\cal C}'$ contour.

To estimate the integrand at $|s|=\Lambda_*$, we focus on simple
tree-level UV-completions in which the $e^+e^-\to e^+e^-$ scattering is mediated
by the exchange of a heavy scalar or vector boson coupling with a strength
$g_{\rm BSM} \simeq 1$. This leads to two distinct cases.
\begin{enumerate}
\item $M(s',0) \sim g^2_{\rm BSM}\frac{s'}{s'-M^2_\mathrm{BSM}}$, which
  corresponds to an $s$-channel scalar or vector exchange. This yields the
  violation
  \be
    \int_{{\cal C}'}\frac{\ud s'}{2\pi i} \frac{M(s',0)}{(s'-2m_e^2)^3} \simeq
     \frac{1}{\Lambda_*^4}\ .
  \ee
\item $M(s',0) \sim {s' g^2_{\rm BSM}}/{M_{\rm BSM}^2}=s'/\Lambda_{\rm BSM}^2$,
  which corresponds to a $t$-channel scalar or vector exchange. This yields the
  violation
  \be
  \int_{{\cal C}'}\frac{\ud s'}{2\pi i} \frac{M(s',0)}{(s'-2m_e^2)^3} \simeq
    \frac{1}{\Lambda_*^2 \Lambda_{\rm BSM}^2}\ .
  \ee
\end{enumerate}

We should emphasize that these simple scenarios do not lead to any
positivity violation. We are solely using them as rough estimates for the
potential size of the boundary term in Eq.~\eqref{eq:disp}, assuming that the
Froissart bound approximately holds below $\Lambda_*$. This shows that, in the
first case, we can use $\Delta$ as an estimate of the scale $\Lambda_*$ at which
the fundamental principles of QFT are violated. In contrast, in the second case,
$\Delta=\sqrt{\Lambda_\mathrm{BSM}\Lambda_*}$. As we have argued that
$\Delta>\Lambda_{\rm BSM}$ (see Section~\ref{sec:violation}), $\Delta$ is thus
lower than the actual scale $\Lambda_*$.
As both $\Delta$ and $\Lambda_{\rm BSM}$ can be inferred from a measurement,
one can estimate $\Lambda_*$ as $\Delta^2/\Lambda_{\rm BSM}$.

The fact that $\Lambda_*$ is either around (first scenario) or above (second
scenario) $\Delta$ is consistent with the assertion of
Ref.~\cite{Distler:2006if}. However, any further determination beyond this rough
estimation is difficult without an apparent characterization of the BSM
nature. But what are the possible scenarios in which positivity may be
violated? There are several possibilities that will be (incompletely) enumerated
in the following.

First, for the UV completions of the SM that are intertwined with the massless
graviton, the $e^+e^-\to e^+e^-$ scattering amplitude contains a $t$-channel
pole with an $s^2$ dependence, and it blows up in the forward limit { 
(see {\it e.g.} ref.~\cite{Bellazzini:2019xts})}. Therefore, the usual
twice-subtracted dispersion relation and the second-order $s$-derivative bound
of Eq.~\eqref{disprelation} cannot be directly used. It is observed
that the standard positivity bounds, which merely require the subtraction
of the infinite $t$-channel pole, can be violated, as shown in some explicit
examples in refs.~\cite{Alberte:2020jsk,LdRST} (see also the work of
ref.~\cite{Bellazzini:2019xts} for the opposite argument). This
violation is conjectured to be suppressed by the quantum gravity scale, which is
usually the Planck scale, although it could be much lower. For example, in ADD
models~\cite{ArkaniHamed:1998rs,ArkaniHamed:1998nn} with two extra dimensions,
the fundamental scale for gravity is around the TeV scale.

If one assumes that the UV theory follows the Regge behavior,
\be
  \Im[M(s\to \infty, t\to 0)] = f(t)\left(\frac{\alpha^{\prime} s}{4}\right)^{2+j(t)},
\ee
where $\alpha'=M_s^{-2}$ is the ``string scale'' of gravity,
one can exactly quantify positivity violation~\cite{Tokuda:2020mlf},
\be\bsp
 \frac{\ud^2 M}{\ud^2 s}& >-\frac{f \alpha^{\prime 2}}{4 \pi}\left[\frac{f^{\prime}}{f j^{\prime}}+\ln \left(\frac{\alpha^{\prime} M_{*}^{2}}{4}\right)\right]\\
 &~~~~~~~~~~~+\frac{f \alpha^{\prime 2}}{4 \pi}\left[\frac{j^{\prime \prime}}{2\left(j^{\prime}\right)^{2}}+O\left(\frac{1}{\alpha^{\prime} M_{*}^{2}}\right)\right] .
\esp\ee
In this expression, the primed quantities refer to a derivative with respect to
$t$, evaluated at $t=0$, and $M_*$ represents the scale at which the Regge
behavior is first observed. In this sense, a test of positivity violation would allow us
to probe the quantum gravity scale and study the implications of low scale
gravity.

Second, one could devise a simple example demonstrating how unitarity
violation leads to positivity violation in the Effective Field Theory (EFT).
One such popular model having an interesting bearing for inflation consists of
the so-called DBI model~\cite{Alishahiha:2004eh}, whose effective
Lagrangian is given by
\be
\label{LDBI}
\mathcal{L}_{\rm EFT}=\varepsilon \Lambda^{4}-\varepsilon \Lambda^{4} \sqrt{1-\frac{\varepsilon(\partial \phi)^{2}}{2\Lambda^{4}} } ,
\ee
with $\varepsilon=1$. Expanding around $(\pd\phi)^2=0$, the leading interaction
term is given by $\varepsilon (\pd\phi)^4/(32\Lambda^4)$ and the corresponding
forward positivity bound is satisfied. In contrast, if we choose instead
$\varepsilon=-1$ in Eq.~\eqref{LDBI}, we obtain the so-called anti-DBI model \cite{Mukhanov:2005bu} that
features positivity violations. To view this from a UV perspective, we recall
that the DBI and anti-DBI models can be derived from a two-field (partial) UV
theory~\cite{Tolley:2009fg}, whose Lagrangian is given by
\be
\label{pUV}
\mathcal{L}_{\rm pUV} = \frac{ (\pd \chi)^2}{2} +  \frac{\varepsilon  e^{\frac{\chi}{M}}(\pd \phi)^2}2 - \Lambda^4\(\cosh \frac{\chi}{M}-1\) ,
\ee
where $\varepsilon$, $\Lambda$, and $M$ are constant. At low energies, the heavy
field $\chi$ is frozen. Neglecting its kinetic term and integrating
it out semi-classically, we obtain the Lagrangian of Eq.~\eqref{LDBI}. From the
Lagrangian of Eq.~\eqref{pUV}, we observe that the $\phi$ field is a ghost (with $\varepsilon=-1$). {Alternatively, we can choose $\epsilon=1$ and send $\Lambda^4$ to $-\Lambda^4$ in Eq.~(\ref{pUV}), which again leads to the anti-DBI model. In this case, $\phi$ is not a ghost, but now $\chi$ is a tachyon \cite{Babichev:2018twg}, and the potential is unbounded from below.} As ghosts {or runaway potentials} lead to some of the worst
instabilities in QFT~\cite{Woodard:2015zca}, positivity violations originating from
this type of UV pathologies appear unlikely.

Then, which of these QFT axiomatic principles is the weakest link? Arguably, it
might be the polynomial boundedness/locality. Indeed, it is widely believed that
gravity is non-local, and the UV completions of general relativity, such as string
theories, violate polynomial boundedness~\cite{Keltner:2015xda}. This is
intimately linked to the observation that black holes are formed in high-energy
scatterings with gravity included, and their horizon radius increases with the
scattering energy. Moreover, there are no local gauge invariant observables in
gravity.

When contemplating UV completions for the SMEFT, one may consider that
general relativity is also an EFT that needs to be UV completed and which might
{\it a priori} be interconnected to the UV theory of the SMEFT. Hence, the SM and
gravity may be (partially) UV completed together, potentially at an
energy scale such as the TeV scale as in models with large extra dimensions.
However, polynomial boundedness is also violated in some innocent looking
(Minkowski space) field theories derived by taking certain low-energy limits of
gravitational theories. For example, the galileon theory consists of a scalar
EFT whose Lagrangian is given by
\be
  {\cal L} = \frac12 \pd^\mu\pi \pd_\mu \pi + \frac{\ai}{\Lambda^3} \pd^\mu\pi \pd_\mu \pi \pd^\rho\pd_\rho \pi + \cdots
\ee
This Lagrangian possesses a generalized shift symmetry when the
galileon field $\pi \to
\pi+c+b^\mu x^\mu$, with $c$ and $b_\mu$ being constants~\cite{Nicolis:2008in}.
Such a setup arises in the decoupling limit of either the DGP braneworld
model~\cite{Dvali:2000hr} or dRGT massive gravity~\cite{deRham:2010kj}. As
the $(\pd^\mu\pi \pd_\mu \pi)^2$ term is forbidden by the generalized shift
symmetry, the 2-to-2 scattering amplitude of this theory does not contain any
$s^2$ term, so that the forward positivity bound is automatically violated.

Discussions on positivity bounds and their implications have recently
restarted in the context of the galileon theory~\cite{Adams:2006sv}. As the
violation is
marginal, adding a softly-breaking mass term for the galileon allows one to
satisfy the forward positivity bounds \cite{deRham:2017avq, Cheung:2016yqr, Bellazzini:2016xrt, Bellazzini:2017fep}. Moreover, at least in some parameter
space region, generalized, $t$-derivative, positivity bounds~\cite{
deRham:2017imi, deRham:2017zjm, deRham:2018qqo} are fulfilled. However,
further generalized positivity bounds exclude the entire
parameter space~\cite{Wang:2020xlt}.

An important feature of the galileon theory along with the DGP model and the dRGT
model is that they embed the so-called Vainshtein mechanism (see
Ref.~\cite{Keltner:2015xda} and the references therein). It has been argued that
theories including the Vainshtein mechanism should not have standard UV
completions whose low-energy EFT satisfies the positivity bounds. Instead, the
high-energy behavior is characterized by a phenomenon called classicalization,
where semi-classical contributions dominate~\cite{Dvali:2010jz,Keltner:2015xda}
{(see ref.~\cite{Vikman:2012bx} for discussions on classicalization in the
anti-DBI model)}.
Examples of those non-standard UV completions also appear in gauge theories, in
the context of chiral perturbation theories~\cite{Aydemir:2012nz}.

Lorentz invariance is also at risk of being violated at high energies. After
all, our intuition of Lorentz invariance comes from low energy and weak gravity
environments. In general relativity, Lorentz invariance still holds in local
inertial frames, but that may just be a prejudice. Hence, Lorentz violating models are
widely discussed in several contexts~\cite{Liberati:2009pf}, including
Horava-Lifshitz gravity~\cite{Horava:2009uw}. In this case, gravity is Lorentz
violating in the UV, so that the theory is potentially renormalizable and flows
to Lorentz-invariant general relativity at low energies (so that it may be
relevant for collider physics).

Finally, one last possibility that could justify a positive $\Delta$ quantity
may be the existence of new states at or not too far above the TeV scale, making
the SMEFT framework invalid. Naively, in such cases, one expects to either
directly produce the new states or observe large deviations in various channels.
Depending on the UV model, it might still be possible that this ``positivity
violation'' is a first indication that BSM physics exists at low scales,
invalidating the SMEFT framework.

In this study, we take an agnostic approach, leaving all these possibilities
open, and, as a first step, focus on the phenomenological feasibility of probing
positivity violation effects. If $\Delta>0$ can be verified experimentally, it
implies that either at least one of the fundamental principles is violated at
the TeV scale or not too far above (so that unconventional new physics is
required, as shown above in this subsection), or that the SMEFT is invalid (which
is also a useful guidance).
In the first case, the logical follow-up requires exploring specific scenarios
under which the violation occurs by using the scale $\Delta$ as an instructive handle
connected to $\Lambda_*$. The precise pin-down of this violation scale in
connection with specific UV models is nevertheless left for future works.

\section{Inferring the existence of new physics states in the UV}
\label{sec:inverse}
The possibility of inferring the existence of new physics states in the UV by
virtue of the positive nature of the dimension-8 Wilson coefficient parameter
space has been recently demonstrated~\cite{Zhang:2020jyn}. It relies on convex
geometry, and some of its basic concepts are listed below.
\begin{itemize}
 \item A {\it convex cone} (or cone) is a subset of a vector space that is
  closed under additions and positive scalar multiplications.
  A {\it salient} cone is a cone that contains no straight line. Thus, if
  $\mc C$ is salient, having both $x\in \mc C$ and $-x\in \mc C$ implies
  $x=0$.
 \item  An {\it extremal ray} of a convex cone $\mc C_0$ is an element
  $x \in\mc C_0$ that is not a sum of two other elements in
  $\mc C_0$. If we can write an extremal ray as $x=y_1+y_2$ with $y_1,y_2\in
  \mc C_0$, we must have $x=\lambda y_1$ or $x=\lambda y_2$,
  $\lambda$ being a real constant. For example, the extremal rays of a
  polyhedral cone are its edges.
 \item The {\it convex hull} of a given set $\mc X$ is the ensemble of all
  convex combinations of points in $\mc X$, where a convex combination is
  defined as a linear combination of points where all the combination
  coefficients are non-negative and add up to 1.
 \item The {\it conical hull} of a given set $\mc X$ is the ensemble of all
  positive linear combinations of elements in $\mc X$, denoted by cone($\mc X$).
  The extremal rays of cone($\mc X$) are a subset of $\mc X$.
 \item The {\it Krein-Milman theorem}~\cite{KM} states that a salient cone $\mc C$ 
  is the convex hull of its extremal rays.
\end{itemize}

We now begin with the second-order $s$-derivative of the forward elastic
scattering $ij\to kl$, which we denote by $M^{ijkl}$ and is related to the UV amplitude via the dispersion
relation~\cite{Zhang:2020jyn}
\begin{flalign}
	M^{ijkl}= \int_{(\epsilon\Lambda)^2}^{\infty}\ud\mu
	{\sum_{Z\text{ in }\mathbf{r}}}'
	\frac{|\braket{{Z}|\mathcal{M}|\mathbf{r}}
	|^2 }{\pi\left( \mu-\frac{1}{2}M^2 \right)^3}
	P_\bfr^{i(j|k|l)}\ .
	\label{eq:2}
\end{flalign}
As more extensively detailed in Ref.~\cite{Zhang:2020jyn}, $\sum'$ denotes a
summation over all possible intermediate $Z\in\bfr$ states along with their
phase space, and $\bfr$ runs through all irreps of the $SO(2)$ rotations around the
forward scattering axis and the gauge symmetries of the SM. Moreover,
$P_\bfr^{ijkl}\equiv \sum_\alpha C^{\bfr,\alpha}_{i,j}(C^{\bfr,\alpha}_{k,l})^*$
represents the projective operators of ${\bf r}$, where $C^{\bfr,\alpha}_{i,j}$
are the Clebsch-Gordan coefficients relevant for the direct sum decomposition of
$\mathbf{r}_i\otimes\mathbf{r}_j$, with $\mathbf{r}_i(\mathbf{r}_j)$ being the
irrep of the particle $i$ ($j$) and with $\alpha$ labeling the states
included in $\mathbf{r}$. The parentheses in $i(j|k|l)$ indicate that the $j,l$ indices
are symmetrized. Finally, $M^2$ is the sum of the four interacting particle
squared masses.

The l.h.s.~of Eq.~\eqref{eq:2} can be expanded in terms of the dimension-8
Wilson coefficients $\vec C^{(8)}$, whereas each $P_\bfr$ projector on the
r.h.s.~can also be written in terms of the Wilson coefficients as
$\vec c^{\ (8)}_{\bfr}$. Positivity arises because all the other factors in
Eq.~\eqref{eq:2} are positive definite, so that
\begin{flalign}
	\vec C^{(8)}\in\mbox{cone}\left( \left\{ \vec c^{\ (8)}_{\bfr} \right\} \right)
	\equiv \mc C\ .
	\label{eq:cone}
\end{flalign}
A nontrivial feature is that $\mc C$ is a salient cone~\cite{Zhang:2020jyn}
whose vertex is at the origin of the Wilson coefficient space so that the cone
does not cover the entire space. Thus, not all possible values for the dimension-8
coefficients are allowed, which leads to the positivity bounds.

As a consequence of the Krein-Milman theorem, this cone $\mc C$ is a convex hull
of its extremal rays, the latter being a subset of $\big\{ \vec c^{\ (8)}_{\bfr}
\big\}$. Geometrically, an element that is close to
some extremal ray $\vec c^{\ (8)}_{\bfr'}$ cannot be decomposed as a positively
weighted sum of other $\vec c^{\ (8)}_{\bfr}$ with $\bfr\neq\bfr'$. This has an
interesting physical consequence: if the measured value of $\vec C^{(8)}$ is
close to an extremal ray, then we can infer that UV states in the $\bfr'$ irrep
must exist and generate the dominant contribution to $\vec C^{(8)}$.
Alternatively, if $\vec C^{(8)}$ is observed to be consistent with 0, then we can
exclude the potential existence of any UV particle, regardless of the model
setup, up to certain scales depending on the precision of the measurement. This
originates from the extremality of the origin in $\mc C$, as the cone is salient.

This new physics inference or exclusion would not be possible at the dimension-6
level due to the absence of any positive nature. For example,
even if $\vec C^{(6)}$ is observed to be consistent with 0, UV particles might
still exist. In this case, the dimension-6 effects would cancel each other out, either
accidentally or consequently to some symmetry~\cite{Gu:2020thj} (or be
suppressed compared with the dimension-8 ones~\cite{Bellazzini:2018paj,
Bellazzini:2017bkb,Liu:2016idz}). In other words, positivity implies both that
the leading BSM effects might not appear at dimension-6, and
that they must not vanish
at dimension-8. Therefore, excluding the presence of dimension-8 effects in observables
is a definitive way to confirm the SM ultimately.

\begin{table}
  \setlength\tabcolsep{6pt}
  \renewcommand{\arraystretch}{1.6}
  \begin{tabular}{ccc|cc}
  \multicolumn{3}{c|}{\mbox{Scalar}}&
  \multicolumn{2}{c}{\mbox{Vector}} \\\hline
  $D \equiv {\bf 2}_{1/2}$& $M_L \equiv {\bf 1}_1$ & $M_R\equiv {\bf 1}_2$&
  $V\equiv {\bf 1}_0$& $V'\equiv {\bf 2}_{-3/2}$
  \end{tabular}
  \caption{New physics degrees of freedom in our UV setup aiming at illustrating
    the strength of the positivity bounds in inferring or excluding the
    existence of new states. The quantum numbers refer to $SU(2)_L$ and $U(1)_Y$,
    respectively.
  \label{tab:type}}
\end{table}

For illustration purposes, a weakly coupled UV theory whose EFT
manifestation is generated by integrating out some heavy states at the tree level is considered.
Nevertheless, the conclusions obtained in the following are valid for loop-level
and non-perturbative cases as well, as the positive nature of the dimension-8
parameter space originates from the dispersion relation of Eq.~\eqref{eq:2} that
holds in general.

We thus extend the SM in the UV by several generations of the
new states shown in Table~\ref{tab:type}, each of them being identified by a
different set of quantum numbers specified by an irrep $\bf r$. These states
couple to the SM electrons through the interaction Lagrangian
\be\bsp
  &\mathcal{L}_{\rm int} =
      g_{Di} \bar L e D_i
    + g_{M_Li} \bar{L}^c
    %\!\cdot\!
    \epsilon L M_{Li}
    + g_{M_Ri} \bar{e}^c e M_{Ri}\\
  &\ +g_{Vi} \Big(\bar{L}\gamma^\mu L + \kappa_i \bar{e}\gamma^\mu e \Big) V_{i\mu}
     + g_{V'i} (\bar{e}^c\gamma^\mu L) V'^\dagger_i\\ &\ + {\rm h.c.},
\esp\label{eq:lagnp}\ee
where the index $i$ is a generation index and
the parameters $\kappa_i$ are arbitrary real numbers. The $g_X$ couplings
correspond to a Dirac-type scalar coupling ($g_D$), Majorana-type scalar
couplings to left-handed ($g_{M_L}$) and right-handed ($g_{M_R}$) fermions, and
vector couplings involving same-chirality ($g_V$) and opposite-chirality
($g_{V'}$) fermions. Under moderate assumptions, these represent all possible
tree-level interactions that can generate dimension-6 four-electron
operators~\cite{deBlas:2017xtg}.\footnote{The fields of Table~\ref{tab:type}
correspond to the $\varphi$, $\mathcal{S}_1$, $\mathcal{S}_2$, $\mathcal{B}$ and
$\mathcal{L}_3$ states introduced in Ref.~\cite{deBlas:2017xtg}. The latter also
includes an $SU(2)_L$ triplet $\mc W$ that we have omitted as we
ignore the $O_5$ operator. As already mentioned in
Section~\ref{sec:ops}, adding the $O_5$ operator has no effect on the subspace
associated with the first four dimension-8 operators relevant for $e^+e^-\to
e^+e^-$ scattering.}

By integrating out these particles, the resulting dimension-8 operator
coefficients are obtained as follows, considering a specific particle species $X$ at a time:
\begin{flalign}
 &\vec{C}_{X}^{(8)}\equiv  
 \sum_i \vec{C}_{Xi}^{(8)}= \sum_i w_{Xi} \vec c^{\ (8)}_X  ,
 \label{eq:3}
\end{flalign}
where one sums over all generations of particles $X$. The ``weights'' $w_{Xi}$
are defined by
\begin{flalign}
	&w_{Xi}=\frac{g_{Xi}^2}{M_{Xi}^{4}}\ge0\ ,
\end{flalign}
with $g_{X_i}$ and $M_{Xi}$ being the mass and coupling of the $i^{\rm th}$
generation of type-$X$ particle respectively. The vectors $\vec c_X^{\ (8)}$ are
constant and are given by
\be\bsp
\vec c^{\ (8)}_D          =&\ (0,0,1,0),\\
\vec c^{\ (8)}_{M_L}      =&\ (0,0,0,-1),\\
\vec c^{\ (8)}_{M_R}      =&\ (-1,0,0,0),\\
\vec c^{\ (8)}_{V'}       =&\ (0,0,-1,2),\\
\vec c^{\ (8)}_{V(\kappa)}=&\ (-\kappa^2/2,-\kappa,0,-1/2).
\esp\label{eq:c82}\ee
Unlike all other particle species, the $V$-type couplings involve a free
parameter $\kappa$, so that different $V$ fields associated with different
$\kappa$ values are considered as different particle species. Summing over all
particle types, the dimension-8 coefficients are given by
\begin{flalign}
	\vec C^{(8)} = \sum_X w_X \vec c^{\ (8)}_X ,
	\label{eq:C8}
\end{flalign}
with $w_X=\sum_i w_{Xi}\ge0$ being the total contribution from each type of
particle. This implies that, for any tree-level UV completions of the SM,
${\vec C}^{(8)}$ is a positively weighted sum of the $\vec c^{\ (8)}_X$ vectors.
Thus, we can define a convex cone $\mc C_1$ that consists of the set of all
possible $\vec C^{(8)}$ vectors that can be generated at the tree level,
\begin{flalign}
	\vec C^{(8)}\in \mc C_1\equiv \mbox{cone}\left( \left\{ \vec c^{\ (8)}_X \right\} \right) ,
	\label{eq:cone1}
\end{flalign}
where $X$ runs through all possible states.

The set of vectors $\{ \vec c^{\ (8)}_X \}$ forms a subset of the
$\{\vec c_\bfr^{\ (8)}\}$ ensemble of vectors appearing in Eq.~\eqref{eq:cone},
with $\bfr$ being the irrep of the $X$ particle species up to a positive
rescaling. Therefore, this results in $\mc C_1\subset\mc C$. Thus, the relation
$\vec C^{(8)}\in\mbox{cone}\big( \big\{ \vec c^{\ (8)}_{\bfr} \big\} \big)$ is
a consequence of the positivity of the weights $w_{Xi}\ge0$, and the
boundary of $\mc C_1$ defines the positivity bounds for the considered
tree-level UV-completion of the SM. In contrast, the boundary of $\mc C$
reflects the bounds that are relevant for any UV completion.

\begin{figure}
 \begin{center}
  \includegraphics[width=.8\linewidth]{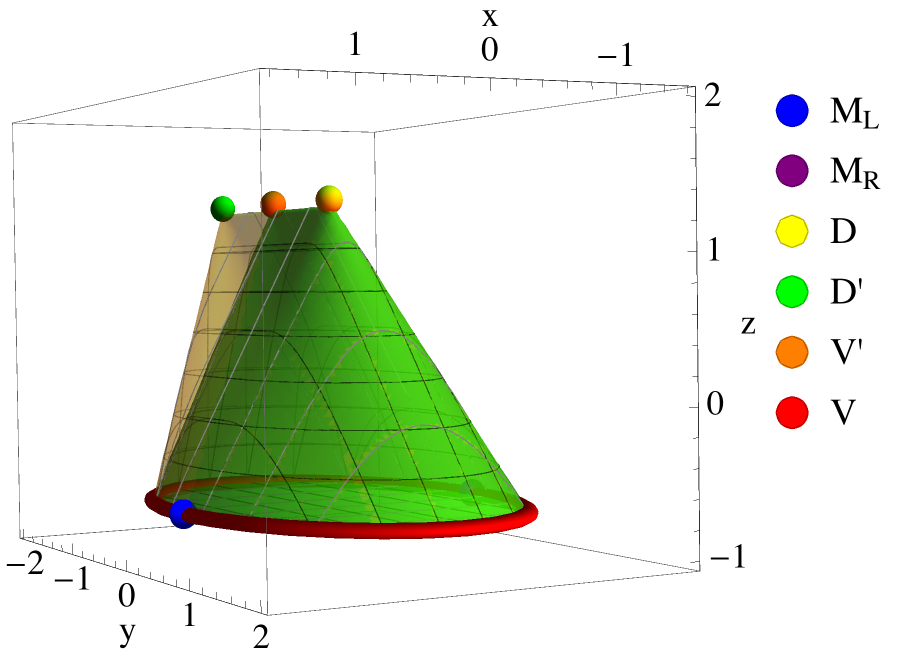}
  \caption{Three-dimensional cross section of the convex cones ${\cal C}$
    (yellow, bigger cone) and ${\cal C}_1$ (green). The cross section is taken
    to be perpendicular to the direction $(1,1,0,1)$. The three axes $x$, $y$,
    and $z$ are defined along the $(1,-1,0,0)$, $(0,0,1,0)$, and $(-1,-1,0,2)$
    directions, respectively.
  \label{fig:d8}}
 \end{center}
\end{figure}

In Figure~\ref{fig:d8}, we present three-dimensional cross sections for both the
$\mc C$ and $\mc C_1$ cones. While $\mc C_1$ has been derived from
Eq.~\eqref{eq:cone1}, $\mc C$ has been obtained as described in
Section~\ref{sec:positivity}, using the elastic scattering of superposed states.
The dimension-8 coefficient vectors $\vec c^{\ (8)}_X$ become points in this
three-dimensional space, although in the case of the $V$ particle species, these
points form a circle as they are continuously parameterized by varying the
$\kappa$ parameter. The cross section of $\mc C_1$ is then the convex hull of
these points. Significant parts of the $\mc C$ and $\mc C_1$ cones coincide,
as $\{\vec c^{\ (8)}_X\}$ is a subset of $\{\vec c^{\ (8)}_{\bfr}\}$, which
serves as a nontrivial check as $\mc C$ is computed with a different approach.

An important observation is that, for tree level UV-completions, all
$\vec c^{\ (8)}_X$ listed in Eq.~\eqref{eq:c82} are extremal, {\it i.e.}, they
cannot be split as positive sums of other elements in $\mc C_1$. Beyond this
tree-level assumption ({\it i.e.}, on the yellow $\mc C$ cone), only one of the
points, $V'$, becomes non-extremal. This is related to the appearance of one
extra irrep, $D'$, so that $V'$ can now be expressed as a positively weighted
sum of $D'$ and $D$. The new irrep $D'$ can be interpreted as a new vector
$V_{D'}$ that couples as a dipole moment so that it cannot be generated by a
tree-level UV completion (and is thus external to the $\mc C_1$ cone).

Let us now assume that the Wilson coefficients can all be determined
experimentally, with the measurement being denoted $\vec C_\mathrm{exp}$. We can thus
write
\begin{flalign}
 \vec C_\mathrm{exp} = \sum_X w_X \vec c^{\ (8)}_X .
\label{eq:exp}\end{flalign}
What can we learn about the particles living in the UV of the theory, and how
important are their effects? In other words, how could we obtain information on
the $w_X$ parameters from the knowledge of the l.h.s.~of the above equation? One may
naively believe that this is not possible as there is an infinite number of ways
to arrange UV particles to satisfy the above equation. Surprisingly, the
positivity nature of the dimension-8 coefficient parameter space allows for
interesting inferences. While in principle, $\vec C_\mathrm{exp}$ can
only be
measured up to some uncertainties, we neglect the latter in this section. We
refer instead to Section~\ref{sec:infer} for a more realistic example including
those uncertainties, in which we demonstrate that this does not prevent us from
using positivity to infer knowledge on the existence of potential BSM
particles and their properties.

We start by considering a measurement of $\vec C_\mathrm{exp}$ that would be
found parallel to a $\vec c^{\ (8)}_{X'}$ vector extremal in $\mc C_1$.
In this case, $\vec C_\mathrm{exp}=\lambda \vec c^{\ (8)}_{X'}$ so that the
only solution to Eq.~\eqref{eq:exp} is
\begin{flalign}
 w_X=\left\{\begin{array}{ll}
  \lambda\quad&\mbox{if }X=X'\ ,\\
  0&\mbox{otherwise} \ .
 \end{array}
\right.
\end{flalign}
This follows from the definition of an extremal ray that cannot be written as a
positively weighted sum of other cone elements. In other words, if the
dimension-8 operators have been generated by particles living in a single
representation, a ``precise'' measurement of the dimension-8 Wilson coefficients
could not only confirm this hypothesis
but also exclude the potential existence
of any other particles without making any assumption on the BSM details,
and thus falsifying other alternative hypotheses. This
feature can be traced back to the fact that all $\vec c_X^{\ (8)}$ live in a
salient cone.

In contrast, the above inference is not possible when one truncates the SMEFT
expansion at the dimension-6 level, as there would always be an infinite number
of positive solutions (with $w_X\ge0$) for Eq.~\eqref{eq:exp}. For example, one
could consider a $D$ scalar and a $V'$ vector with arbitrarily large
contributions that cancel each other out completely, or similarly an $M_L$ scalar
and a $V$ vector with $\kappa=0$. This originates from the fact that the allowed
values for the dimension-6 Wilson coefficients do not live within a salient
cone, and hence, there are always several ways to organize them to reproduce any measured
coefficient value. Therefore, an interpretation at the dimension-6 level can
only be achieved under specific model assumptions and thus fits within a
top-down study.

More generally, if $\vec C_\mathrm{exp}$ is not extremal, one can still set upper
limits on the weights $w_X$ by starting from
\be
 \vec C(\lambda)\equiv \vec C_\mathrm{exp}-\lambda \vec c_{X'}
   =\sum_{X\neq X'} w_X\vec c_X+(w_{X'}-\lambda)\vec c_{X'} .
\label{eq:clam}\ee
Since $\vec C(0)\in \mc C_1$ and $\lim_{\lambda\to+\infty}\vec C(\lambda) \notin
\mc C_1$ (by the definition of a salient cone), there exists a maximum value
$\lambda_\mathrm{max}$ below which we have $\vec C(\lambda)\in \mc C_1$. This
provides an upper bound for $w_{X'}$, as if $w_{X'}>\lambda_\mathrm{max}$, then
we can find a $\lambda$ value such that $w_{X'}>\lambda>\lambda_\mathrm{max}$
and for which $\vec C(\lambda)\notin \mc C_1$ (as $\lambda>
\lambda_\mathrm{max}$), yielding a contradiction. Physically, this indicates that, if
we remove some $X'$ contribution from $\vec C_\mathrm{exp}$, the remaining
vector still consists of a positively weighted sum that should satisfy
tree-level positivity by belonging to $\mc C_1$. Therefore, the largest $X'$
contribution that could be removed from $\vec C_\mathrm{exp}$ without spoiling
these bounds provides an upper bound on $w_{X'}$. This can then be iteratively
used to set upper limits on the existence of all types of particles. Again, such
an inference is not possible at the dimension-6 level, as there is no equivalent
to $\mc C_1$.

Setting an upper limit on $w_X$ is important, as this limit applies to not only
the total contribution from all particles of a given type but also
each individual generation of a particle of this type. While this is evident
for $X\neq V$, as
\begin{flalign}
 0\le\frac{g_{Xi}^2}{M_{Xi}^4}\le  \sum_i\frac{g_{Xi}^2}{M_{Xi}^4}=w_{X} ,
\end{flalign}
this is also true for $X=V$, as all $\vec c_{V(\kappa)}^{\ (8)}$ live on a
circular cone.

A similar reasoning can be achieved beyond the tree level by replacing the $\mc C_1$
cone by $\mc C$. Hence, upper limits can be set on the existence of states in all
possible irreps $\bf r$ and on an individual generation of particles
lying in this irrep. Moreover, this includes both one-particle and
multi-particle states (which yield loop-level generated coefficients), as their
contributions are always individually positive.

In summary, we have shown so far that, in contrast to the dimension-6 case, a
measurement of the dimension-8 Wilson coefficients would allow us to rule out
or at least place a lower bound on the mass scale of each individual particle
of a given type $X$ without any model assumption. If a deviation from the SM is
observed, these universal bounds narrow down the possible range of UV-complete
BSM models that should be considered. On the contrary, if no deviation is
observed, then model-independent exclusion limits on the BSM states can be set,
at least up to certain scales depending on the precision of the measurement.
This last point is crucial as a test of the SM. If no significant deviation from
the SM is observed at future colliders, a global fit of the dimension-6 Wilson
coefficients would only allow to set limits on the dimension-6 contributions
without being able to further exclude the possibility that BSM exists in a way
yielding the suppression of any dimension-6 effect (by virtue of cancellations
or symmetry reasons). Hence, such a fit would not be sufficient to confirm the
SM. In contrast, a global fit of the coefficients of operators ranging up to
dimension-8 would allow for not only for the extraction of limits on the
coefficients, but, more importantly, also the exclusion of the
existence of BSM states, thus confirming the SM.

The illustration presented in this section is based on the assumption of
weakly-coupled UV completions. However, the conclusions hold in general, as
positivity implies that any UV completion of the SM must lead to some
non-vanishing dimension-8 effects. This should further motivate the study of
dimension-8 operators through precision physics in the future.

\begin{table*}
  \setlength\tabcolsep{6pt}
  \renewcommand{\arraystretch}{1.4}
  \begin{tabular}{cc|cccc}
    \multirow{2}{*}{Scenario} & Beam polarization &
    \multicolumn{4}{|c} {Runs (luminosity @ energy), [ab$^{-1}$]~@~[GeV]}\\
    & $P(e^-,e^+)$
    & \hspace*{1cm}1 \hspace*{1cm} 
    & \hspace*{1cm}2 \hspace*{1cm} 
    & \hspace*{1cm}3 \hspace*{1cm} 
    & \hspace*{1cm}4 \hspace*{1cm} \\\hline&&&&&\\[-.4cm]
    CEPC   & None & 2.6@161 & 5.6@240 & &\\[.2cm]
    FCC-ee & None & 10@161 & 5@240 & 0.2@350 & 1.5@365\\[.2cm]
    \multirow{2}{*}{ILC-500}
      & $(-80\%, 30\%)$ &0.9@250 & 0.135@350 & 1.6@500&\\
      & $(80\%, -30\%)$ &0.9@250 & 0.045@350 & 1.6@500&\\[.2cm]
    \multirow{2}{*}{ILC-1000}
      & $(-80\%, 30\%)$ &0.9@250 & 0.135@350 & 1.6@500& 1.25@1000 \\
      & $(80\%, -30\%)$ &0.9@250 & 0.045@350 & 1.6@500& 1.25@1000 \\[.2cm]
    \multirow{2}{*}{CLIC}
      & $(-80\%, 0\%)$ & 0.5@380 & 2@1500 & 4@3000 &\\
      & $(80\%, 0\%)$ &0.5@380 & 0.5@1500 & 1@3000 &
  \end{tabular}
  \caption{Different future collider operation runs considered in this study,
    presented together with the associated center-of-mass energy, expected
    luminosity, and beam polarization setup (if relevant). \label{tab:runs}}
\end{table*}

On different grounds, we indicate that it is also possible to set some lower
limits on a particular weight $w_{X'}$. To this end, we introduce the
convex cone $\mc H_{X'}$ defined by all $\vec c_X$ vectors different from
$\vec c_{X'}$, using instead the opposite of the latter as a last element to
define the cone,
\be
  \mc H_{X'}=
    \mbox{cone}\left( \left\{ \vec c_{X\neq X'}, -\vec c_{X'}\right\}\right).
\label{eq:upperl}\ee
In the case where $\vec C_\mathrm{exp}\notin \mc H_{X'}$, the lower limit on
$w_{X'}$ is given by the minimum $\lambda$ value such that
$\vec C_\mathrm{exp}-\lambda\vec c_{X'}\in \mc H_{X'}$. However, this also
applies at the dimension-6 level.

\section{Collider analysis}
\label{sec:collider}
In this section, we pioneer a realistic study of the feasibility of positivity
tests at future colliders and investigate the possibility of inferring or
excluding the existence of new physics in the UV. However, a more accurate determination
of the impact of dimension-8 contributions at future colliders (and the
estimation of their reach in the corresponding Wilson coefficient parameter
space) is beyond the scope of this paper. Hence, we have
performed several simplifications in our analysis. We first restrict
ourselves to parton-level simulations, and omit any higher-order
corrections and initial-state radiation effect. Second, we assume an ideal
detector and thus ignore any reconstruction and experimental effect.

We utilize {\sc FeynRules}~\cite{Alloul:2013bka} to generate a UFO
model~\cite{Degrande:2011ua} including all the operators introduced in
Section~\ref{sec:ops} so that we could simulate $e^+e^-\to e^+e^-$ scattering
with {\sc MadGraph5\_aMC@NLO}~\cite{Alwall:2014hca}. We analyze the resulting
parton-level events with {\sc MadAnalysis}~5~\cite{Conte:2012fm,Conte:2014zja,
Conte:2018vmg} which is also used to define an appropriate fiducial volume. The
latter embeds a selection on the final-state lepton transverse momentum of
$p_T>5$~GeV and on their pseudorapidity of $|\eta|<5$. We then split the phase
space into 25 bins in $\cos\theta$, with $\theta$ being the lepton scattering
angle. Moreover, we discard the most forward bin, as it corresponds to the bin
where the SM contribution blows up.

\begin{figure*}
  \begin{center}
    \includegraphics[width=.8\linewidth]{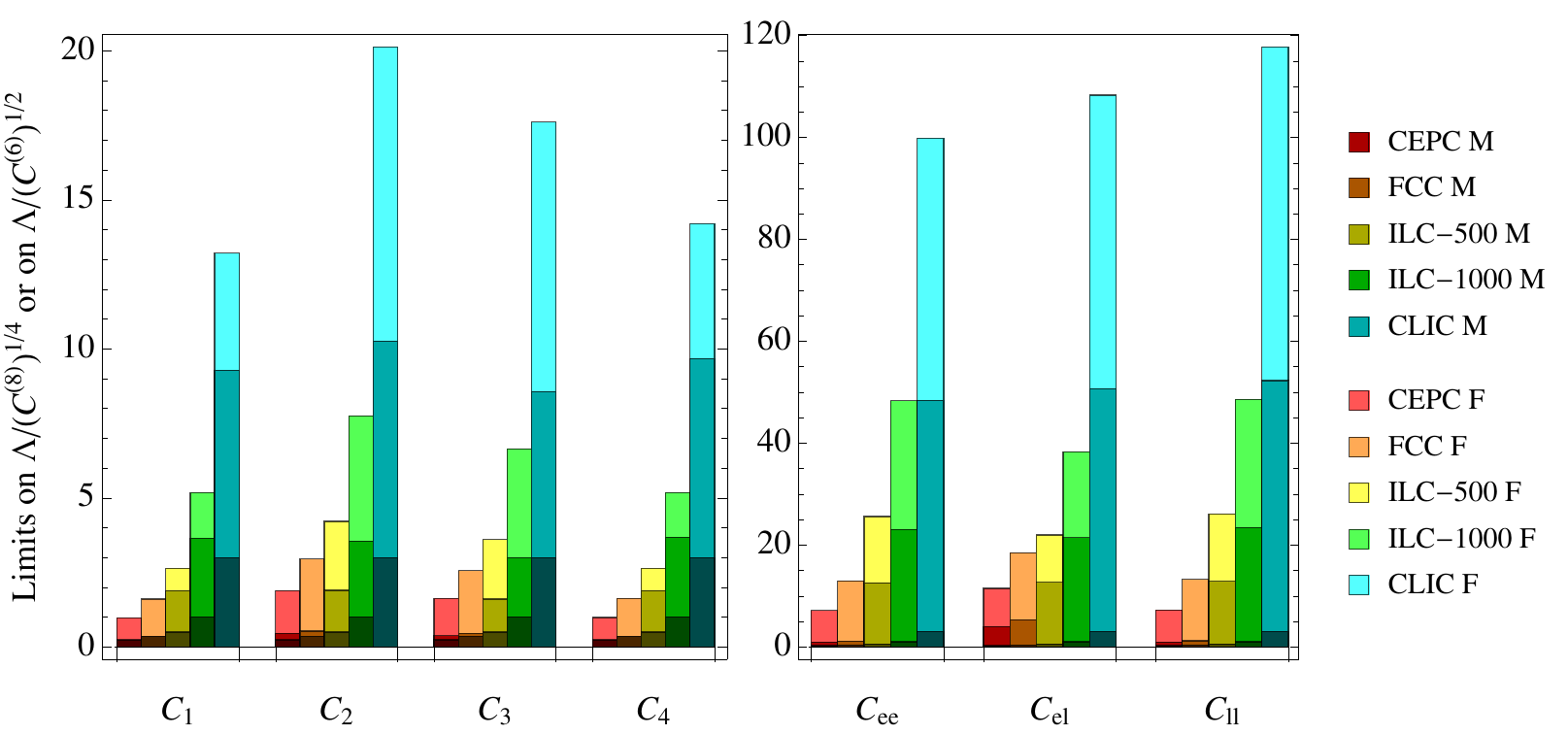}
    \caption{Limits on the new physics characterization scale $\Lambda_c$ (in
      TeV) for the various considered future lepton colliders. ``M'' denotes
      marginalized limits (all other coefficients being floating) whereas ``F''
      denotes individual limits (all other coefficients being vanishing). In
      addition, we represent by the darkest color the largest center-of-mass
      energy of each collider project.\label{fig:1}}
  \end{center}
\end{figure*}

Given that $e^+e^-\to e^+e^-$ cross section
measurements at LEP2 reached a precision of approximately $2\%$, we assume that the
systematic uncertainties could be controlled at the $1\%$ level for each
$\cos\theta$ bin. In addition, we include statistical uncertainties that we
estimate as $\sqrt{N}$, where $N$ is the projected number of events in a given bin.
We have finally verified the consistency of the simulation results with
analytical calculations.

We have considered several future lepton collider projects, mostly following the
setups presented in Ref.~\cite{deBlas:2019wgy}. However, we have omitted any
operation run at the $Z$-pole as the cross section is dominated by the
$Z$-resonance
contribution instead of any potential four-fermion operator effect. In addition,
for the ILC case, we focus on a possible upgrade at a center-of-mass energy
of 1~TeV~\cite{Asner:2013psa}. We refer to Table~\ref{tab:runs} for details on
the center-of-mass energies, luminosities, and beam polarization options of all
collider configurations studied in this paper.

\subsection{Future lepton collider sensitivity to four-electron operators}

The $e^+e^-\to e^+e^-$ differential cross section can be parameterized as a
polynomial in the higher-dimensional operator coefficients, as shown in
Eq.~\eqref{eq:poly}. In the following, we denote by $\vec C$ the entire set of
coefficients,
\be
   \vec C =(C_{ee},C_{el},C_{ll},C_1,C_2,C_3,C_4)\ ,
\ee
and the effective cutoff scale $\Lambda$ is set to 1~TeV, unless specified
otherwise. To evaluate the constraining power and sensitivity of the future
collider under consideration to the higher-dimensional operators introduced in
Section~\ref{sec:ops}, we built a $\chi^2$ function,
\be
 \chi^2\left(\vec C,\vec C_0\right)\ ,
\ee
in which we assume that a would-be observation $\vec C$ agrees with the theoretical
predictions associated with some reference hypothesis $\vec C= \vec C_0$. The
allowed range for $\vec C$ at some confidence level can then be determined by
the $\vec C$ values for which $\chi^2\le \chi^2_c$, where $\chi^2_c$ represents
the critical $\chi^2$ value allowing to reach an agreement with $\vec C_0$ at
the required confidence level. For the remainder of this paper, all the presented limits
have been evaluated at $2\sigma$. 

In the case where the would-be observations would agree with the SM, limits on
the four-electron Wilson coefficients can be set by enforcing $\chi^2(\vec C,0)
\le \chi^2_c$. These limits, cast under a $C\in [C_{\rm min},C_{\rm max}]$ form
for each coefficient,
reflect the sensitivity of the future lepton colliders to the considered
operators. They can be further converted into the new physics characterization
scale $\Lambda_c$ that represents the BSM scale that is reachable at the various
colliders. We define $\Lambda_c$ by
\be\bsp
  \mbox{dimension-6:}\quad &
    \Lambda_c\equiv\frac{\Lambda}{\sqrt{\frac{C_{\rm max}-C_{\rm min}}{2}}}\ ,\\
  \mbox{dimension-8:}\quad &
    \Lambda_c\equiv\frac{\Lambda}{\sqrt[4]{\frac{C_{\rm max}-C_{\rm min}}{2}}}\ ,
\esp\ee
in the dimension-6 and dimension-8 cases, respectively.

These scales are depicted in Figure~\ref{fig:1}. For each collider and each
specific coefficient ({\it i.e.}, in each column), the lighter color represents
the individual limit obtained when all other
coefficients are enforced to vanish, whereas the darker color represents the
marginalized limit obtained where all other coefficients are left floating. For
an easy comparison of the strengths of the various machines, we show through
the darkest color the largest collider center-of-mass energy reachable for each
collider.

In the case of the dimension-6 operator coefficients, we observe that all future
lepton colliders are sensitive to very large new physics scales (given in terms
of $\Lambda_c$). Those scales are indeed typically approximately 1 or 2
orders of magnitude larger than the collider center-of-mass energy. This is
more or less expected, given that LEP2 already reached a sensitivity of a few
TeV on those operators~\cite{Han:2004az}.

For the dimension-8 operators, the individual sensitivities range from
$\mathcal{O}(1)$ (CEPC) to $\mathcal{O}(10)$ (CLIC) TeV. They are roughly a
factor of 5 larger than the collider center-of-mass energy, which also indicates
that the EFT approach is robust in the considered context. However,
the more reliable limits are those obtained through marginalized bounds.
They are reduced by a factor of a few when compared with the individual limits.
Nevertheless, for all scenarios, the corresponding $\Lambda_c$ is sufficiently higher
than the collider energy, except for the $O_1$ and $O_4$ operators in the
CEPC and FCC-ee cases. Here, the $\Lambda_c$ scale is observed to be slightly lower than
the highest expected center-of-mass energy (see the discussion below). In
general, the EFT validity is thus not an issue for $\mathcal{O}(1)$ BSM
couplings, and the dimension-8 effects can be identified even in the presence of
dimension-6 operators.

For all considered circular colliders, the marginalized limits for the operators $O_1$,
$O_4$, $O_{ee}$, and $O_{ll}$ are much weaker than their corresponding
individual limits. This originates from an accidental degeneracy between the
linear-level contributions of the $LLLL$ and $RRRR$ types of operators. For a
slightly larger electroweak mixing angle such that $\sin^2\theta_W=0.25$, the
$Zee$ coupling in the SM would be of a purely axial-vector nature, yielding
identical $e^+_Le^-_L\to e^+_Le^-_L$ and $e^+_Re^-_R\to e^+_Re^-_R$ cross sections
(including the photon and $Z$-boson exchanges, as well as their interference).
In this hypothetical case, the $O_{ee}$/$O_{ll}$ and $O_1$/$O_4$ contributions
cannot be distinguished, unless beam polarization is used (so that the linear
collider sensitivity is much stronger). In practice,
$\sin^2\theta_W = 0.234$ (from $m_Z=91.1876$ GeV and $\alpha=1/127.9$), and hence,
this degeneracy is not exact. However, this leads to an almost flat direction in
the Wilson coefficient parameter space, or equivalently to large differences
between the individual and marginalized limits for the $O_1$, $O_4$,
$O_{ee}$, and $O_{ll}$ operators.

As our fit is at the quadratic level for the dimension-6 operator
coefficients, $O_{ee}$ and $O_{ll}$ are thus essentially constrained by their
quadratic contributions. On the other hand, $O_1$ and $O_4$ are only included at
the linear level so that some caution is required when interpreting the
corresponding limits. It is observed that the numerical simulations are not
reliable in this case, as statistical fluctuations may artificially lift the
degeneracy. Therefore, we have employed our analytical computations for those
two operators and the two circular collider cases. We have additionally verified
that, using $m_Z$, $G_F$, and $m_W$ as electroweak input parameters (yielding
$\sin^2\theta_W=0.223$), the marginalized limits on $O_1$ and $O_4$ are only
impacted at the level of approximately 10\%, all other limits being stable.

Finally, to estimate the error due to the SMEFT truncation at
${\cal O}(\Lambda^{-4})$ in Eq.~\eqref{eq:poly}, we have assessed the impact of
the next order contributions, namely the interferences between the dimension-6
and dimension-8 operators and the quadratic contributions in the dimension-8
operators. The resulting changes in Figure~\ref{fig:1} are quite mild. The
limits at circular colliders are modified by less than 10\%, whereas
those associated with the linear colliders are negligibly affected. Therefore
our truncation at ${\cal O}(\Lambda^{-4})$ is reliable, and any higher-order
contributions will be ignored in the remainder of this paper.

\subsection{Testing positivity at future lepton colliders}
\label{sec:test}
We now assume that some would-be observation at future colliders in $e^+e^-\to
e^+e^-$ scattering data is consistent with a coefficient value hypothesis
$\vec C_0$. We aim at investigating what we could learn, from this measurement,
about the potential amount of positivity violation $\Delta^{-1}$.

We start from the fact that, for any given
$\vec C_0$, a measurement indicates that the true coefficient vector $\vec C$ of the
theory is constrained by $\chi^2({ \vec C,\vec C_0})<\chi^2_c$ at some
confidence level. We can thus deduce a confidence interval for the amount of
positivity violation,
\be
   \Delta^{-1}\in [\Delta^{-1}_\mathrm{low},\Delta^{-1}_\mathrm{high}]\ ,
\ee
with
\be\bsp
  \Delta^{-1}_\mathrm{low} = &\ \min_{\chi^2({ \vec C,\vec C_0})\le\chi^2_c}
    \left(\frac{\delta(\vec C)^\frac{1}{4}}{\Lambda}\right),\\
  \Delta^{-1}_\mathrm{high} = &\ \max_{\chi^2({ \vec C,\vec C_0})\le\chi^2_c}
    \left(\frac{\delta(\vec C)^\frac{1}{4}}{\Lambda}\right).
\esp\ee
We focus on $\Delta^{-1}_\mathrm{low}$, which is a conservative
estimate of $\Delta^{-1}$, so that we could conclude about the existence of some
positivity violation if $\Delta_\mathrm{low}^{-1}>0$.

We consider four benchmarks that differ by the $\vec C_0^{(8)}=(C_1,C_2,C_3,C_4)$
choice,
\be\bsp
  B_1:\ & \vec{C}_0^{(8)}=(0,0,3,1.2),\\
  B_2:\ & \vec{C}_0^{(8)}=(0,0.3,0.2,0),\\
  B_3:\ & \vec{C}_0^{(8)}=(0,0.015,0.015,0),\\
  B_4:\ & \vec{C}_0^{(8)}=(0,0,0.0006,0.00015).
\esp\label{eq:bench}\ee
The corresponding amount of positivity violation $\Delta^{-1}$ is given in
Table~\ref{tab:BM}, together with the $\Delta^{-1}_\mathrm{low}$ values that
could be reached at each considered collider scenario when assuming that the
measurements are consistent with the $B_i$ hypothesis. Those results have been
estimated by marginalizing over all dimension-6 four-electron operators, so that
they can be taken as conservative.

\begin{table}
  \setlength\tabcolsep{5pt}
  \renewcommand{\arraystretch}{1.6}
  \begin{tabular}{l|c|ccccc}
   & \multirow{2}{*}{$\Delta^{-1}$} & \multicolumn{5}{|c}{$\Delta^{-1}_{\rm low}$}\\
   && CEPC & FCC-ee & ILC-500 & ILC-1000 & CLIC \\\hline
   $B_1$ & 1.48 & 0 & 0.86 & 1.45 & 1.47 & 1.48\\
   $B_2$ & 0.74 & 0 & 0 & 0.66 & 0.73 & 0.74\\
   $B_3$ & 0.35 & 0 & 0 & 0 & 0.29 & 0.35\\
   $B_4$ & 0.16 & 0 & 0 & 0 & 0 & 0.10
  \end{tabular}
  \caption{Amount of positivity violation $\Delta^{-1}$ associated with each
    considered benchmark and the corresponding $\Delta^{-1}_{\rm low}$
    bound that could be obtained at each collider scenario. The results are all
    given in TeV$^{-1}$ and at the 95\% confidence level. \label{tab:BM} }
\end{table}

The four points have been chosen to illustrate an increasing sensitivity to
positivity violations at the five collider scenarios under
consideration. The $B_1$ setup corresponds to a violation arising at a scale of
approximately 700 GeV, which can be observed at the $2\sigma$ level by all studied
lepton colliders except for the CEPC. The $B_2$ point allows for the
observation of a positivity violation at scales of approximately 1.4 TeV, which can only
be observed at future linear colliders. Finally, the $B_3$ and $B_4$ benchmarks
induce positivity violation scales that go up to 2.9 and 6.4 TeV,
respectively, to which only the ILC with an energy upgrade at 1~TeV and CLIC are
expected to be sensitive.

\begin{figure}
  \begin{center}
    \hspace*{-1cm}Benchmark 3\\
    \includegraphics[width=\columnwidth]{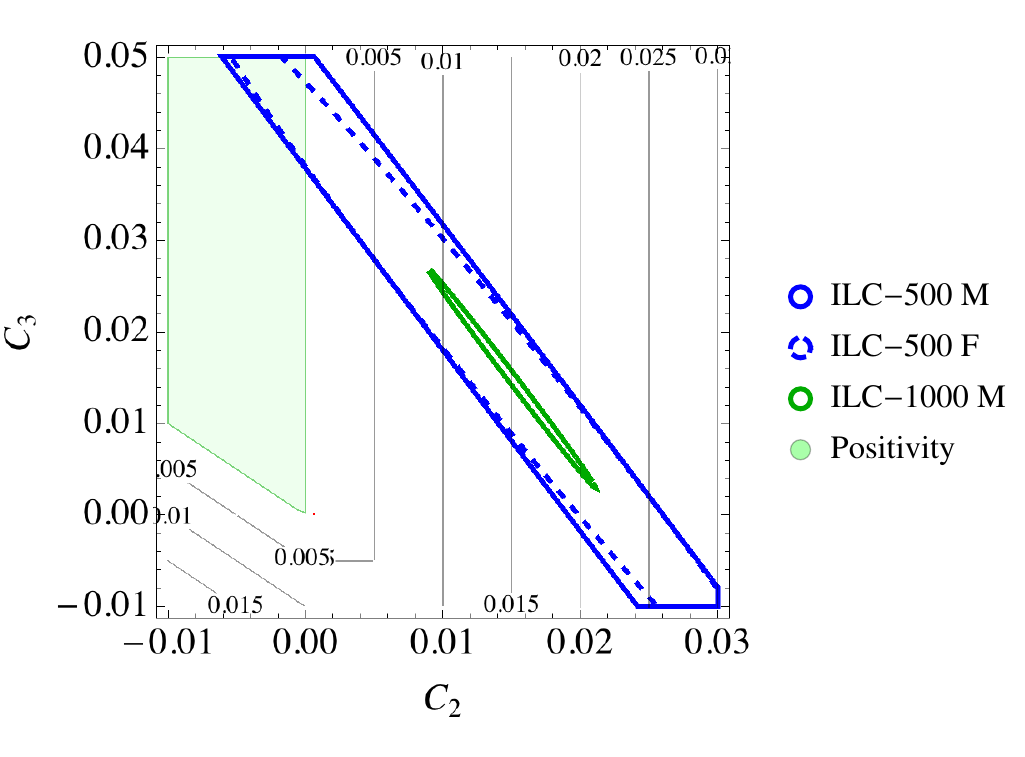}\\
    \hspace*{-1cm}Benchmark 4\\
    \includegraphics[width=\columnwidth]{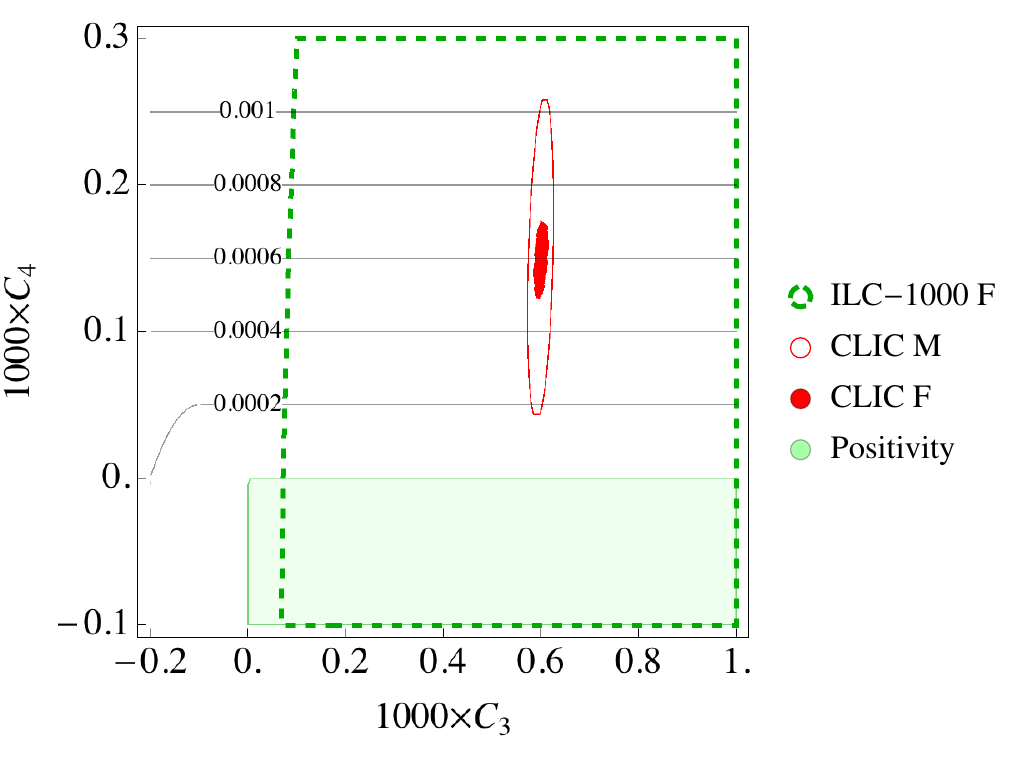}
    \caption{Bounds on the pairs of dimension-8 operator coefficients that are
    relevant for the benchmark points $B_3$ (upper) and $B_4$ (lower). In each
    case, the other
    two dimension-8 coefficients are set to 0, as in the benchmark scenario
    definitions from Eq.~\eqref{eq:bench}, whereas the dimension-6 coefficients
    are either marginalized over (M, solid) or fixed to 0 (F, dashed). We show
    the $2\sigma$ expectations for the reach of different future lepton
    collider projects as colored contours, and the light green area represents
    the parameter space region allowed by the positivity bounds. Outside this
    area, the gray isolines depict the dependence of the $\delta(\vec C)=
    ({1\ \mathrm{TeV}}/{\Delta})^4$ quantity, which is thus used as an
    estimate for the amount of positivity violation, on the coefficients.
     \label{fig:BM34}}
  \end{center}
\end{figure}

To obtain a more intuitive picture, we present a slice of the dimension-8
Wilson coefficient parameter space for the two benchmarks $B_3$ and $B_4$ in
Figure~\ref{fig:BM34}. In the two subfigures, we depict by a light green area
the parameter space region allowed by the positivity bounds. Moreover, we
indicate through light gray contours the amount of positivity violation arising
in the remainder of the parameter space, using the dimensionless parameter
$\delta(\vec C^{(8)}) =({1\ \mathrm{TeV}}/{\Delta})^4$. Consistent with the
definition of these two benchmarks in Eq.~\eqref{eq:bench}, the other two
dimension-8 coefficients ($C_1$ and $C_4$ for the $B_3$ scenario and $C_1$ and
$C_2$ for the $B_4$ scenario) are taken as vanishing. The limits that could be
imposed from measurements at various lepton colliders are given by solid and
dashed contours. They respectively correspond to a derivation including a
marginalization over the dimension-6 Wilson coefficients (labeled by ``M'') or after
fixing them to zero (labeled by ``F'').

We observed differences between the solid and dashed contours, which indicates that
there are
correlations between the impacts of the dimension-6 and dimension-8
operators. Nevertheless, even after marginalizing the dimension-6 coefficients, a
sensitivity to the dimension-8 operators remains, as illustrated in the case of
the ILC-1000 collider (for the $B_3$ benchmark) and CLIC (for the $B_4$
benchmark). The entire $2\sigma$ contours indeed lie outside the positivity
area so that positivity violation could be confirmed,
regardless of the existence of any dimension-6 effect. On different grounds, notably,
the fact that the marginalized limits do not
overlap with the positivity area does not guarantee a potential confirmation of
$\Delta^{-1}_\mathrm{low}>0$, as we only focus here on a slice of the full
dimension-8 coefficient space.

It is evident that a large $\Delta^{-1}$ value has a better chance to be confirmed
experimentally, but this also depends on the actual values of all dimension-8
coefficients. Any given amount of violation may indeed indicate different
regions of the parameter space, with some of them being phenomenologically easier to
detect than others. An interesting question would be as follows: how large should the
amount of violation $\Delta^{-1}$ be for a collider to have a significant chance
to confirm it?

\begin{figure*}
 \begin{center}
  \includegraphics[width=.98\columnwidth]{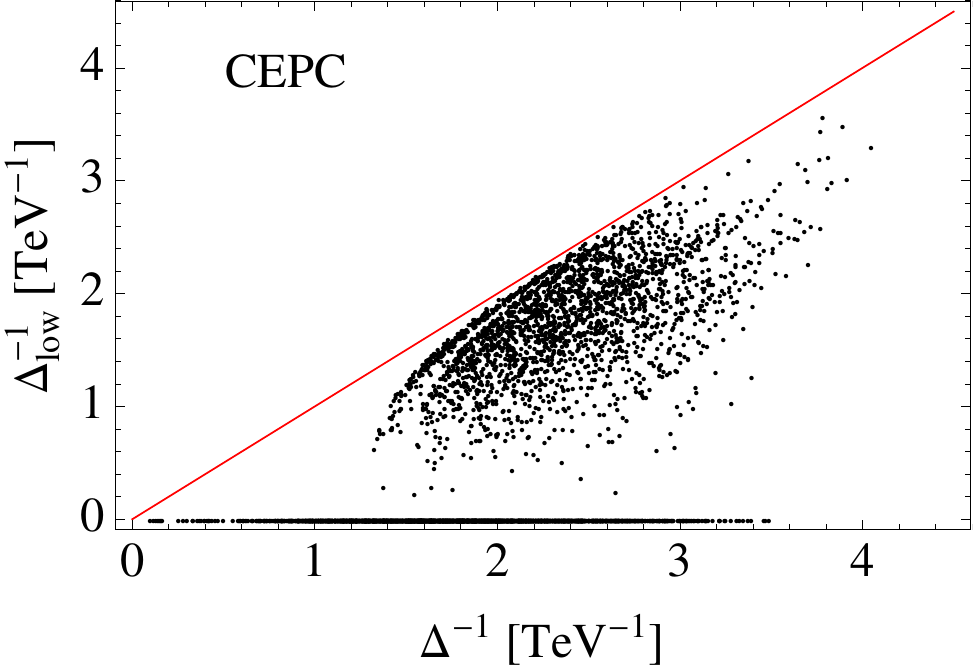}
  \includegraphics[width=.98\columnwidth]{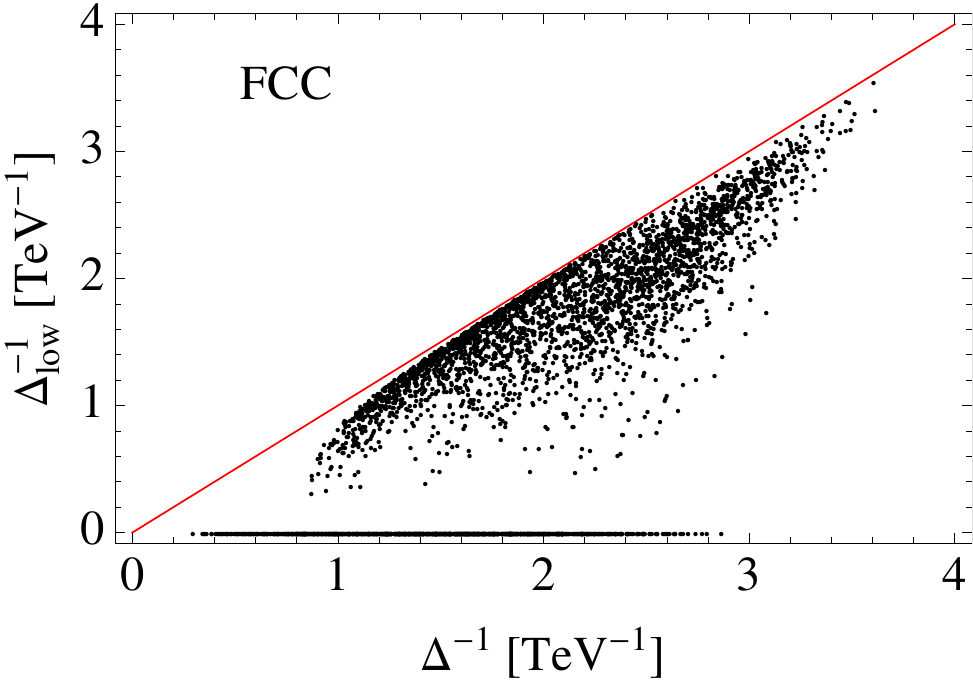}\\[.2cm]
  \includegraphics[width=.98\columnwidth]{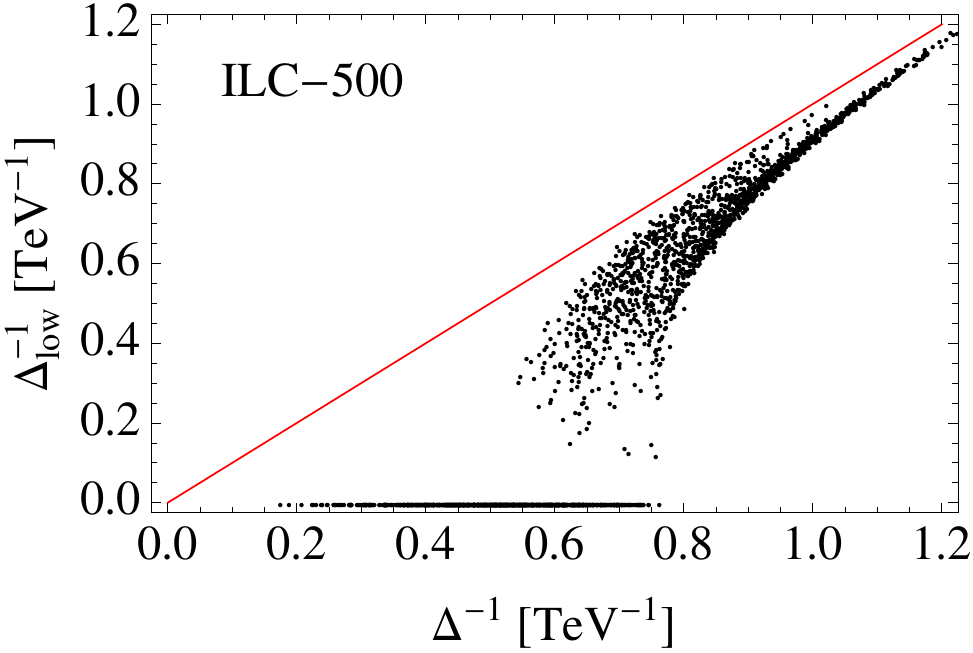}
  \includegraphics[width=.98\columnwidth]{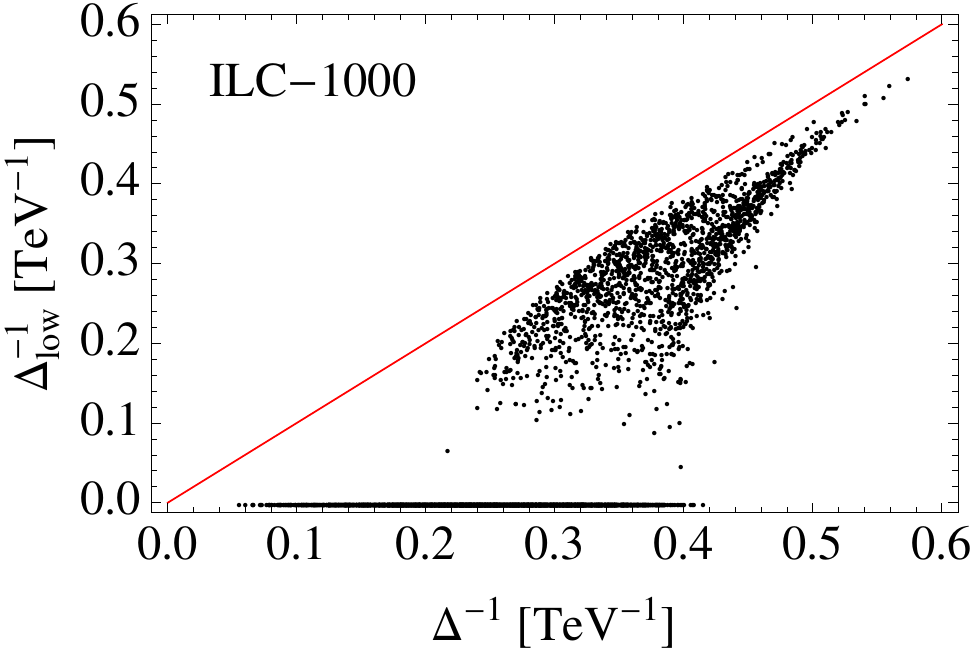}\\[.2cm]
  \includegraphics[width=.98\columnwidth]{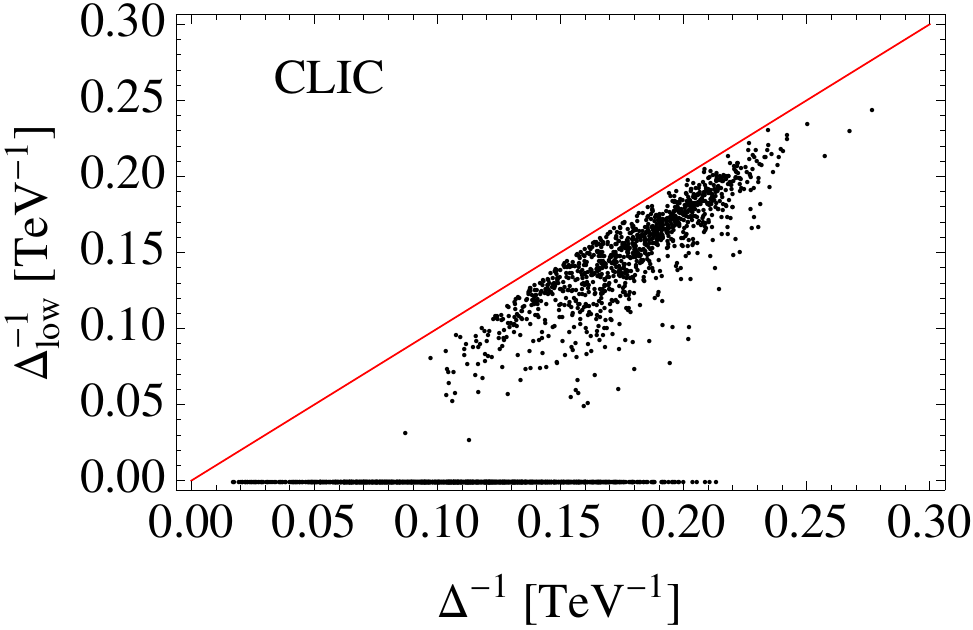}
  \caption{Correlations between the amount of positivity violation
    ($\Delta^{-1}$) and the maximal sensititivity that can be reached at a
    given future lepton collider ($\Delta^{-1}_{\rm low}$). We
    present results for the CEPC (upper left), FCC-ee (upper right), ILC
    (center), and CLIC (lower) colliders, and each represented point has been
    obtained through a Monte Carlo sampling of the dimension-6 and dimension-8
    Wilson coefficient parameter space.\label{fig:CCLC}}
 \end{center}
\end{figure*}

A quantitative and accurate answer is difficult to provide, due to the quartic
nature of our $\chi^2$ fit and the discontinuous nature of the $\delta(\vec C)$
function. We present below a tentative answer by sampling the dimension-6 and
dimension-8 parameter space with a Monte Carlo method and assessing, for each
sampled configuration, the $(\Delta^{-1}, \Delta^{-1}_\mathrm{low})$ values.
After restricting the values of the dimension-6 coefficients to the order
of 0.1~TeV$^{-1}$ so that they are roughly consistent with LEP-2
constraints~\cite{Han:2004az}, we show our results in Figure~\ref{fig:CCLC}.
This figure shows the correlation between the violation scale $\Delta^{-1}$ and
its $2\sigma$ lower bound $\Delta^{-1}_\mathrm{low}$ as obtained by would-be
measurements at the CEPC, FCC-ee, ILC-500, ILC-1000, and CLIC colliders. In other
words, the results indicate the experimental sensitivity to some positivity
violation scale $\Delta$ at a given collider.

Two particularly important
features can be extracted.
First, the smallest $\Delta^{-1}$ value that corresponds to a nonzero
$\Delta_\mathrm{low}^{-1}$ value defines the minimum amount of positivity
violation that is, in principle, observable at each collider. This corresponds to
scales of 0.7, 1.2, 1.8, 4.2, and 11 TeV for the CEPC, FCC-ee, ILC-500, ILC-1000,
and CLIC colliders, respectively. These scales are much higher than the
corresponding (highest) expected center-of-mass energy, and thus, the positivity tests
are observed to be phenomenologically feasible for all five machines. Moreover, a
positivity violation that should occur at or below these scales thus has a
chance to be detected. Second, the largest $\Delta^{-1}$ value associated with a
zero $\Delta_{\rm low}^{-1}$ value corresponds to the minimum guaranteed
observable amount of positivity violation, regardless of the actual
coefficient values. This corresponds to scales of 0.3, 0.36, 1.3,
2.4, and 4.8 TeV for the above five colliders, respectively. These scales are slightly higher
than, but comparable to, the corresponding collider energies. Equivalently, a
violation occurring at these scales yields a guaranteed $2\sigma$ observation.

Those possible evaluations of $\Delta$ for $e^+e^-\to e^+e^-$ scattering at
various colliders serve as a proof of concept for how well collider physics can
be used as a novel means to probe the fundamental principles of QFT in a
model-independent way. As we have argued in Section \ref{sec:interpretation},
the scale $\Delta$ could be viewed as a rough estimate for the scale $\Lambda_*$
(with $\Lambda_*\gtrsim \Delta$) connected to the violation of the fundamental
QFT principles. The results obtained in the present section demonstrate that
scenarios for which $\Delta$ lies in a range of $\mathcal{O}(1)-\mathcal{O}(10)$
TeV have a chance to be confirmed.

Therefore, we conclude that future $e^+e^-$
colliders will be able to test the fundamental principles of QFT up to a scale
of the order of 10~TeV or even beyond, depending on the exact nature of BSM
physics. Moreover, if positivity violation is observed, then unconventional
model building approaches, such as those discussed in
Section~\ref{sec:interpretation}, will be necessary. The measurement of the
corresponding $\Delta$ value will, in this case, provide an important guidance.
However, the actual connection between $\Delta$ and the
scale $\Lambda_*$ at which the QFT core principles are violated
should be studied on a model-by-model basis.

\subsection{Inferring/excluding states in the UV}
\label{sec:infer}

In Section~\ref{sec:inverse}, we have argued that the positive nature of the
dimension-8 parameter space can be used to infer or exclude the possible
existence of new physics states in the UV, independent of the nature of the
new physics model. In this section, this argument is demonstrated with realistic
examples.

We first consider an extension of the SM where the new physics sector of the
theory solely includes a $D$-type scalar (see Table~\ref{tab:type}). For an
illustrative benchmark scenario, we fix its mass to 2~TeV, its coupling $g_D$
defined in Eq.~\eqref{eq:lagnp} to 0.8, and focus on the ILC-1000 collider.
Integrating this heavy field out generates two higher-dimensional operators:
one of dimension-6 and one of dimension-8. The associated Wilson coefficients
can be parameterized in terms of the $\vec C_0^{(6)}$ and $\vec C_0^{(8)}$
vectors of Eq.~\eqref{eq:Wilvec},
\be
  \vec C_0^{(6)} = (0,-0.08, 0)\ , \qquad
  \vec C_0^{(8)} = (0,0,0.04,0)\ .
\label{eq:c068}\ee

The $\chi^2$-fit introduced in the previous section allows for the
identification of the coefficient space region that would be reachable from
$e^+e^-\to e^+e^-$ measurements at the ILC-1000, assuming a $\vec C_0$ theory
hypothesis. Such a region is defined by the $\vec C$ values yielding
$\chi^2(\vec C,\vec C_0)\le \chi_c$ so that one can extract marginalized
limits, at the 95\% confidence level, on all coefficients,
\be\bsp
 &C_{ee}=0\pm0.0024, \qquad \ C_{el}=-0.08\pm0.0035, \\
 &\hspace*{2.3cm} C_{ll}=0\pm0.0023,\\
 &C_1=0\pm0.0074,    \qquad\  \ C_2=0\pm0.0077,\\
 &C_3=0.04\pm0.020,  \qquad C_4=0\pm0.0071 .
\esp\ee

As already mentioned above, the interpretation of these results cannot be
model-independent at the dimension-6 level. For
example, assuming that the SM is only supplemented by a $D$-type scalar
generating the $O_{el}$ operator, one would then obtain as a bound on the new
physics mass scale
\be
  M_D/g_D\in [2.45,2.56]\ \mbox{TeV} .
\label{eq:dim6bound1}\ee
On the other hand, if we assume that the SM is extended by both a $D$
scalar and a $V'$ vector (coupling with a strength $g_{V'}$ as in
Eq.~\eqref{eq:lagnp}), then we can only conclude that
\be
  \frac{g_D^2}{2M_D^2}-\frac{g_{V'}^2}{M_{V'}^2}=0.08\pm0.0035\ \mathrm{TeV}^{-2}.
\label{eq:dim6bound2}\ee
Moreover, it is impossible to disentangle the individual contributions from
each particle type, and the situation only worsens by making the model more
complex due to other ways by which similar cancellations may occur.

This shows that conclusions can only be drawn under very specific BSM
assumptions, as we do not know {\it a priori} the exact particle content of the
theory.
Thus, this kind of interpretation is only practical for top-down studies of
specific UV models, as for example the one carried out in Ref.~\cite{Dawson:2020oco}.

One might naively believe that this lack of information is due to the fact that, at the
dimension-6 level, we only measure three coefficients, whereas the number of possible
UV states is infinite (as there can be several $V$ particles with different
$\kappa_i$ couplings). Consequently, including dimension-8 operators as well
would not help significantly. On the contrary, this is not the case. As we have argued in
Section~\ref{sec:inverse}, the positive nature of the dimension-8 space allows
us to set an upper limit on the total contribution (or weights $w_X$) of any
given type of particles $X$.

For example, for a particle species $X'$, we can determine the $\lambda_\mathrm{max}$
value defined by
\be
  \lambda_\mathrm{max} \equiv \max_\lambda \Big[
  \vec C_\mathrm{exp}^{(8)}- \lambda \vec c_{X'}^{\ (8)}\in \mc C\Big] ,
\ee
where $\vec C_\mathrm{exp}^{(8)}$ denotes the projected measurements of the Wilson
coefficients at some collider, the vector $\vec c^{\ (8)}_{X'}$ is given,
for any specific particle type, by Eq.~\eqref{eq:c82}, and $\mc C$ consists of
the cone generated from the entire ensemble of $\vec c^{\ (8)}_X$ vectors. Therefore, the
$\lambda_\mathrm{max}\vec c^{\ (8)}_{X'}$ quantity indicates an upper
bound on the contribution to the dimension-8 coefficients that can arise from
any set of $X'$ states.

This has a simple interpretation at the tree level. If we remove from
$\vec C_\mathrm{exp}$ the contribution of all BSM $X'$ states, then the
remaining quantity is still a positively weighted sum of the contributions of
particles from all types different from $X'$. Consequently, this should fall
within $\mc C$. Therefore, the largest amount that can be removed from
$\vec C_\mathrm{exp}$ in the $\vec c^{\ (8)}_{X'}$ direction
without leaving the cone $\mc C$ provides the upper bound on the
total contribution of $X'$ states to $\vec C_0$. Moreover, this still holds
beyond the tree level, as both the existence of the cone $\mc C$ and the reasoning
provided in Section~\ref{sec:inverse} are valid for all orders.

However, the above statement assumes that $\vec C_\mathrm{exp}^{(8)}$ is ideally
determined, without any uncertainty. In practice, this is not the case,
so that one could question the impact of the experimental
uncertainties. In our example, the measurement suggests that
$\vec C^{(8)}$ falls in the dimension-8 coefficient parameter space region
defined by $\chi^2(\vec C,\vec C_0)\le \chi_c^2$. Therefore, the experimental uncertainties
can be naturally accounted for by evaluating the maximum of all
$\lambda_\mathrm{max}$ derived for all allowed $\vec C^{(8)}$ values in this
region,
\be
  \lambda_\mathrm{max} \equiv \max_\lambda \Big[
    \vec C^{(8)}- \lambda \vec c_{X'}^{\ (8)}\in \mc C;\
    \chi^2\big(\vec C, \vec C_0\big)\le \chi^2_c
    \Big] \ ,
\label{eq:minimize}\ee
where $\vec C\equiv \{\vec C^{(6)}, \vec C^{(8)}\}$ and
$\vec C_0\equiv \{\vec C_0^{(6)}, \vec C_0^{(8)}\}$. Thus, we have
also incorporated the dimension-6 coefficients that are marginalized over.

Let us now apply this general procedure to the benchmark scenario defined in
Eq.~\eqref{eq:c068}. As the coefficients $\vec C^{(6)}_0$ and $\vec C^{(8)}_0$
are known, the $\chi^2$ function can be built from the projected measurement.
Eq.~\eqref{eq:minimize} therefore expresses a constrained optimization problem.
For example, for a $M_V$ new physics vector state with $\kappa=1$ (as defined in
Table~\ref{tab:type} and Eq.~\eqref{eq:lagnp} so that its
couplings to the electron are of a vector-like nature), we find
$\lambda_\mathrm{max}=0.0054$. Thus, the $M_V$ contribution to $\vec C^{(8)}$ is
constrained to be less than approximately 10\% in magnitude. In terms of the new physics
mass and coupling, this is given by
\be
  \frac{M_{V}}{\sqrt{g_{V}}}\ge 3.7\ \mbox{TeV} .
\label{eq:dim8boundMV}\ee
We emphasize that such a bound is different from the dimension-6 case
expressed in Eqs.~\eqref{eq:dim6bound1} and \eqref{eq:dim6bound2}. The
constraint of Eq.~\eqref{eq:dim8boundMV} is indeed universal and excludes any
possible $V$-type particle featuring a coupling to electrons of strength
$\kappa=1$ with properties violating the bound, without relying on any specific
model assumptions.

\begin{figure}
  \begin{center}
    \includegraphics[width=.8\linewidth]{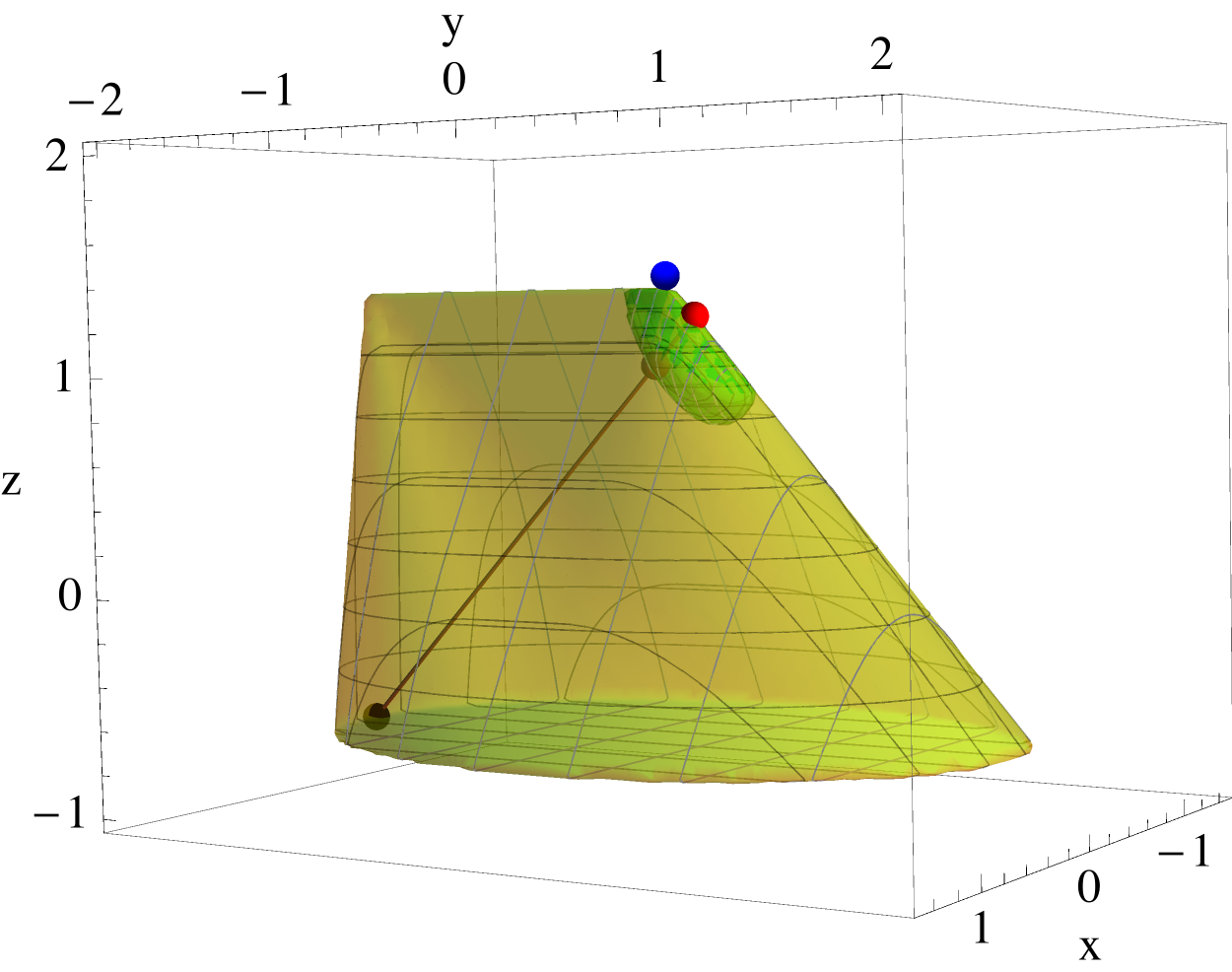}
    \caption{Three-dimensional cross section of the dimension-8 convex cone
      $\mc C$ (yellow). This cross section is extracted as in
      Figure~\ref{fig:d8}. The experimental bounds, marginalized over the
      dimension-6 coefficients, are displayed by the green region.
      The blue, brown, red, and black dots represent the $\vec C^{(8)}_0$,
      $\vec C_{\rm max}^{(8)}$, $\vec C_{\rm max}^{(8)}-\lambda_{\rm max}\vec
      c_{V}^{\ (8)}$, and $\vec c^{\ (8)}_{V}$ vectors, respectively. We refer to
      the text for more details.}
  \label{fig:L}
    \end{center}
\end{figure}

For an intuitive understanding of this universal bound, we
investigate what happens when $\lambda$ varies toward $\lambda =
\lambda_\mathrm{max}$ geometrically. To this end, we identify two points in the
dimension-8 coefficient space: $\vec C_{\rm max}^{(8)}$ and $\vec
C_{\rm max}^{(8)}-\lambda_{\rm max}\vec c_{V}^{\ (8)}$. The former corresponds
to a saturation of the experimental bounds, whereas the latter corresponds to a
saturation of the
positivity bounds. This is represented in Figure~\ref{fig:L}, in which we
present a slice of the dimension-8 coefficient space that we have extracted
in a similar manner to Figure~\ref{fig:d8}. The green region is obtained from the
experimental would-be measurements (after marginalizing over the dimension-6
coefficients), the yellow area consists of a three-dimensional cross section of
the positivity cone ${\mc C}$, whereas $\vec C_{\rm max}^{(8)}$ and $\vec C_{\rm
max}^{(8)} - \lambda_{\rm max} \vec c_{V}^{\ (8)}$ are shown as a brown dot and
red dot, respectively. We additionally indicate by a blue dot and black dot the
true value of the Wilson coefficients $\vec C^{(8)}_0$ and the new physics
contributions, $\vec c^{\ (8)}_{V}$, that we would like to constrain,
respectively.

The black, brown, and red dots lie on a straight line, which stems from the fact
that $\vec C_{\rm max}^{(8)}$ is a positively weighted sum of
$\vec C_{\rm max}^{(8)}-\lambda\vec c_{V}^{\ (8)}$ and $\vec c_{V}^{\ (8)}$.
This figure further illustrates that the maximization process yielding the
$\lambda_{\rm max}$ quantity corresponds to estimating the
largest possible distance between a point consistent with the experimental data
and a point allowed by the positivity bounds in a given direction
$\vec c^{\ (8)}_{V}$. Considering instead a $\lambda$
value larger than $\lambda_{\rm max}$ would mean to rely either on a setup
experimentally excluded, or on a theory violating positivity. As these two
options are evidently excluded, $\lambda$ cannot be larger than
$\lambda_{\rm max}$.

\begin{table}
  \setlength\tabcolsep{5pt}
  \renewcommand{\arraystretch}{1.6}
  \begin{tabular}{cc|cc}
    $X$ & $\vec{c}_{X}^{\ (8)}$ & $\lambda_\mathrm{max}$ & $M_X/\sqrt{g_X}$\\
    \hline
    $M_L$                  & $(0,0,0,-1)$       & 0.0067 & $\ge 3.5$~TeV \\
    $M_R$                  & $(-1,0,0,0)$       & 0.0069 & $\ge 3.5$~TeV \\
    $V$ (with $\kappa=1$)  & $(-1/2,-1,0,-1/2)$ & 0.0055 & $\ge 3.7$~TeV \\
    $V$ (with $\kappa=-1$) & $(-1/2,1,0,-1/2)$  & 0.0116 & $\ge 3.0$~TeV \\
    $V'$                   & $(0,-1,2,0)$       & 0.0109 & $\ge 3.1$~TeV \\
  \end{tabular}
  \caption{Universal bounds imposed with positivity on the particle species $X$,
    and for a BSM setup in which the SM is extended by a $D$-type scalar. The
    bounds, expressed in terms of the $\lambda_{\rm max}$ quantity of
    Eq.~\eqref{eq:minimize} (third column), are also translated in terms of the
    new physics masses and couplings (fourth column). We additionally recall the
    values of the different $\vec{c}_{X}^{\ (8)}$ vectors (second column).
   \label{tab:bounds}}
\end{table}

The same class of universal bounds can be set on all types of potential states
extending the SM in the UV. The Wilson coefficients relevant for our benchmark
assumption $\vec C_0$ of Eq.~\eqref{eq:c068} are generated by integrating
a scalar $D$-field out. However, the approach is most effective for all other
particle types in Table~\ref{tab:type}. We present the results for the $M_L$,
$M_R$, $V$ with $\kappa=\pm1$ ({\it i.e.}, corresponding to a vector and
axial-vector coupling to the electron field) and $V'$ states in
Table~\ref{tab:bounds}. This allows for the derivation, for each particle type,
of the $\lambda_\mathrm{max}\vec c^{\ (8)}_X$ quantity that depicts the maximum
possible contribution from $X$-type particles to the generated dimension-8
operators $\vec C_0^{(8)} = (0,0,0.04,0)$. In most cases, bounds can be set at
the $10-20\%$ level. This corresponds, after converting the results in
terms of BSM scales, to roughly an exclusion of UV particles lighter than
approximately $3-4$~TeV.

In contrast, for $D$-type scalars, we obtain a much larger
$\lambda_\mathrm{max}$ value,
\be
  \lambda_\mathrm{max}\approx 0.049\ .
\ee
Using Eq.~\eqref{eq:upperl}, we could additionally infer a lower limit related
to the existence of all types of particles. Such a limit is observed to be always
0, except for $D$-type particles for which we find
\be
  \lambda_\mathrm{min}\approx 0.011\ .
\ee
All these limits suggest that $D$-type particles should exist in viable
UV completions, whereas all other types of particles are severely constrained.
Such a result is not surprising and is consistent with our benchmark
assumption. Such a conclusion thus provides the first steps to obtaining an
answer to the ``inverse problem''. We emphasize again that the
obtained bounds are universal, in the sense that they apply without any
assumption on the nature of the UV physics. Thus, they are useful to exclude BSM
models and guide the model builders adopting a bottom-up approach.

Perhaps a more practically useful example in the light of the current LHC
results is the SM case itself, where $\vec C_0=0$. In this context, the bounds
on the dimension-6 Wilson coefficients are conventionally interpreted as bounds
on the BSM mass scales and couplings, such as
\begin{flalign}
  \frac{M_{D}}{g_{D}}\ge 16\mbox{ TeV}, \quad
  \frac{M_{M_L}}{g_{M_L}}\ge 17\mbox{ TeV} ,
    \ \ldots
\end{flalign}
Whilst these are rather strong bounds, they can only be derived under very
specific UV-model assumptions in which the SM is supplemented by a single
heavy particle at a time. Therefore, they can easily be relaxed by considering
more complicated models so that this one-particle interpretation is only useful
for a top-down approach imposing a clear BSM setup in the UV. However, from the
perspective of testing the SM, this is still insufficient, as one can never
neither exclude the existence of any given class of BSM states, nor confirm the
SM itself. There are indeed models whose dimension-6 coefficients feature
cancellations or are suppressed for various reasons.

\begin{table}
  \setlength\tabcolsep{5pt}
  \renewcommand{\arraystretch}{1.6}
  \begin{tabular}{c|c|cc}
    $X$ & $M_X/g_X$ & $\lambda_\mathrm{max}$ & $M_X/\sqrt{g_X}$\\\hline
    $D$                    &  $\ge 16$~TeV & 0.0076 & $\ge 3.4$~TeV \\
    $M_L$                  &  $\ge 17$~TeV & 0.0053 & $\ge 3.7$~TeV \\
    $M_R$                  &  $\ge 17$~TeV & 0.0054 & $\ge 3.7$~TeV \\
    $V'$                   &  $\ge 23$~TeV & 0.0056 & $\ge 3.7$~TeV \\
    $V$ (with $\kappa=1$)  &  $\ge 28$~TeV & 0.0041 & $\ge 4.0$~TeV \\
    $V$ (with $\kappa=-1$) &  $\ge 21$~TeV & 0.0041 & $\ge 4.0$~TeV \\
  \end{tabular}
  \caption{Bounds imposed on the existence of a particle species $X$,
    assuming the absence of BSM physics. The bounds are provided as limits on
    model-dependent one-particle extensions of the SM extracted from a fit of
    the dimension-6 coefficients (second column), as well as universal and
    model-independent bounds on a given particle species $X$ derived from
    positivity. The latter are expressed in terms of the
    $\lambda_{\rm max}$ quantity of Eq.~\eqref{eq:minimize} (third column) and
    in terms of the new physics masses and couplings (fourth column).
   \label{tab:smbounds}}
\end{table}

Dimension-8 positivity instead provides a chance to exclude the
potential existence of any class of UV states universally in a bottom-up way, hence without
any specific model assumptions. Focusing on the ILC-1000 collider, we extract
bounds by using the same approach as in the previous example and present the
results in Table~\ref{tab:smbounds}, together with the model-dependent
one-particle dimension-6 bounds. As expected, the universal bounds extracted
from the positive nature of the dimension-8 coefficient parameter space are
weaker than the one-particle dimension-6 bounds. However, they are still much
higher than the collider energy of 1~TeV. Assuming that no deviations
from the SM would be observed at the ILC-1000, these bounds will allow for the
exclusion of any UV model featuring all types of states up to certain scales.
In contrast to the dimension-6 case, it will not be possible to remove those
constraints by adding or arranging the properties of other states, 
unless the latter is done in a way that violates
the positivity bounds.

In summary, studying dimension-8 operators in the context of future
high-precision machines will pave the way for universal exclusions of entire
classes of BSM theories by forbidding the existence of specific types of
new states in the UV. Moreover, null BSM search results will eventually allow
for the confirmation of the SM at scales much higher than the collider energies,
providing the dimension-8 operators a special role in the precision test of the
SM.

\section{Summary}
\label{sec:summary}

In this study, we have investigated the positivity features of the dimension-8
four-electron operators. We have, in particular, derived the complete set of
positivity bounds for the $e^4D^2$ class of operators using the elastic
scatterings of states with arbitrary superpositions. 
Therefore, we have reproduced, by using a different approach, the results
of Ref.~\cite{Zhang:2020jyn},
the latter having been derived using the extremal representation of dimension-8
positive convex cones.
We have then investigated two phenomenological aspects of four-electron
positivity at future lepton colliders.

The first is the possibility of probing
positivity violations, which, if present, will revolutionize our
understanding of the fundamental pillars of QFT or the $S$-matrix theory.
Practically, it would provide very important information for model
building. We have proposed a model-independent quantification of the amount of
violation and discussed the implications of positivity violations in terms of
breaking the axiomatic properties of QFT and paving the way for non-standard UV
completions of the SM. We have observed that positivity violation at scales of
order 1--10~TeV can potentially be probed by all future lepton collider projects
currently discussed within the high-energy physics community. This includes the
CEPC, FCC-ee, ILC (with a possible energy upgrade), and CLIC colliders. At each
collider, positivity violation effects can be probed up to scales of about a factor
of a few higher than the highest expected collider center-of-mass energy,
regardless of the presence of any four-electron dimension-6 operators. This
suggests an important and novel avenue for testing the fundamental principles of
QFT at future lepton colliders. If their violation is observed, a more
tailored description for the violated effect may be designed, and results
could be improved through further model-specific studies.

The second aspect of our study is that, if the fundamental principles of QFT hold in
the UV, the positive nature of the dimension-8 coefficient space allows us
to infer the existence of new states at the UV scales directly and characterize
their quantum numbers from measurements. Conversely, it allows for the exclusion
of certain particles up to scales depending on the future measurement precision.
This originates from several concepts in convex geometry and can be
achieved without any BSM model assumption, in contrast to the conventional
SMEFT interpretation truncated at the dimension-6 level that always requires
a specific model assumption. Therefore, while a dimension-6-level approach
is useful mostly for a top-down investigation of any given model, the
dimension-8-level one is as important, as it provides the possibility of
setting model-independent bounds on certain types of particles in the UV, or
in other words, solving the inverse problem. We have demonstrated this point
with realistic examples, using projected measurements at the ILC with an energy
upgrade at 1~TeV. In particular, if no deviation from the SM is observed, we
have explicitly shown how the existence of any type of heavy particles up to scales
much higher than ILC energies can be excluded, regardless of the UV model setup,
thanks to the requirement that any UV completion of the SM has to satisfy
positivity. This underlines a major difference between dimension-6 and
dimension-8 operators (or the $s$ and $s^2$ term in the amplitude). While one
can design UV-complete models that result in vanishing dimension-6 operators
due to accidental cancellations or for symmetry reasons (which makes it
impossible to exclude reliably the presence of specific states in the UV), this
does not hold at the dimension-8 level. By virtue of the positivity bounds, the
dimension-8 operators are not allowed to vanish if the theory features extra
states in the UV.

Consequently, it is crucial to plan a comprehensive study of the
dimension-8 operator effects, not only at the theoretical level but also at
the experimental level.

\bigskip
\textbf{Note added:} After this paper appeared on the arXiv and was sent
to \textit{Chin.Phys.C}, Ref.~\cite{Remmen:2020uze} appeared on the arXiv.  Its
authors studied ``explicit positivity bounds on dimension-six fermionic
operators in the SMEFT''. Their result contrast with our observation in
Section~\ref{sec:inverse}, where by explicitly integrating out all particle
species in Table~\ref{tab:type}, we observed that any measured dimension-6
coefficient values could yield an infinite number of UV completions. Therefore, no bound
exists. This difference from the results of
Ref.~\cite{Remmen:2020uze} arises from neglecting the existence of new vector
bosons in a UV completion.

\acknowledgements
We would like to thank Claudia de Rham, Andrew J.~Tolley, and Alexander Vikman
for useful comments on the manuscript.  CZ would like to thank Jiayin Gu for
helpful discussions related to collider analysis.  CZ is supported by IHEP
under the Contract No.~Y7515540U1, and by
National Natural Science Foundation of China (NSFC) under grant No.~12035008.
SYZ acknowledges support from the starting grants from University of Science
and Technology of China under grant No.~KY2030000089 and GG2030040375, and is
also supported by NSFC (12075233, 11947301, 12047502) and by the Fundamental Research Funds for the Central Universities (No.~WK2030000036). This work has
been supported by the FCPPL France China Particle Physics Laboratory of the
IN2P3/CNRS.

\appendix

\section{Positivity bounds from elastic scatterings}
\label{sec:a1}
Here, we present details about how to obtain positivity bounds from the
elastic scattering of two superposed states $f_1$ and $f_2$ defined by
\be
  f_{1,2}\equiv \epsilon_{1,2}^i F^i\ \  \text{with}\ \
  F^i\equiv \left( e_R,e_L,\nu_L,\bar e_R,\bar e_L,\bar \nu_L \right)\ ,
\ee
where $\epsilon_{1,2}$ are arbitrary complex six-vectors and the index $i$ is
summed over. We need to ``scan'' the full projective space of these vectors to
exhaust all possible bounds and identify the most constraining ones.

First, we show that mixings between fields with different chiralities do not
provide any new bounds. To this end, we consider a forward scattering $f_1f_2\to
f_1f_2$, where $f_{1,2}$ consist of admixtures of left-handed fields $f_L^i$ and
right-handed fields $f_R^i$,
\begin{flalign}
	f_{1,2}\equiv a_{1,2}^i f_L^i + b_{1,2}^i f_R^i\ .
\end{flalign}
The amplitude can be decomposed as
{\small
\be\bsp
  & M(f_1f_2\to f_1f_2) = \\
    &\ \ \ \
       a^i_1a^{k\dagger}_1 a^j_2a^{l\dagger}_2 M(f_L^if_L^j \!\to\! f_L^kf_L^l)
      \!+\! b^i_1b^{k\dagger}_1 b^j_2b^{l\dagger}_2 M(f_R^if_R^j \!\to\! f_R^kf_R^l)\\
    &\ + a^i_1a^{k\dagger}_1 b^j_2b^{l\dagger}_2 M(f_L^if_R^j \!\to\! f_L^kf_R^l)
      \!+\! b^i_1b^{k\dagger}_1 a^j_2a^{l\dagger}_2 M(f_R^if_L^j \!\to\! f_R^kf_L^l)\\
    &\ + a^i_1b^{k\dagger}_1 a^j_2b^{l\dagger}_2 M(f_L^if_L^j \!\to\! f_R^kf_R^l)
      \!+\! b^i_1a^{k\dagger}_1 b^j_2a^{l\dagger}_2 M(f_R^if_R^j \!\to\! f_L^kf_L^l)\\
    &\ + a^i_1b^{k\dagger}_1 b^j_2a^{l\dagger}_2 M(f_L^if_R^j \!\to\! f_R^kf_L^l)
      \!+\! b^i_1a^{k\dagger}_1 a^j_2b^{l\dagger}_2 M(f_R^if_L^j \!\to\! f_L^kf_R^l),
\esp\ee
}where once again, any repeated index is summed over. In this expression, the
penultimate row vanishes because all the considered operators conserve chirality.
Therefore, the involved fermions cannot flip chirality without a mass factor, and
these contributions do not contribute to the positivity bounds obtained
by performing a second-order $s$-derivative of the amplitude. Moreover, the last
row vanishes because of the conservation of the angular momentum in the forward
limit. Thus, only the first two rows are nonzero, and each remaining contribution
is individually elastic. They can then be considered one by one, and they correspond
to the four forward and elastic scattering matrix elements,
\begin{flalign}
	M(f_{1,R}f_{2,R} \to f_{1,R}f_{2,R}) \label{eq:ampsRR} ,
	\\
	M(f_{1,L}f_{2,L} \to f_{1,L}f_{2,L}) \label{eq:ampsLL},
	\\
	M(f_{1,R}f_{2,L} \to f_{1,R}f_{2,L}) \label{eq:ampsRL},
	\\
	M(f_{1,L}f_{2,R} \to f_{1,L}f_{2,R}) \label{eq:ampsLR},
\end{flalign}
where $f_{n,L}\equiv a^n_1 f_L^i$ and $f_{n,R}\equiv b^n_1 f_R^i$ ($n=1,2$)
represent the superpositions of left-handed fermions and right-handed fermions only,
respectively. The positivity bounds are then fully encoded in the scattering
amplitudes of these fields, which have a definite chirality. Such a feature is
typical of four-fermion scattering in the SMEFT. In contrast, it does not hold
for vector boson scattering, as there is no associated chirality conservation
law in the SMEFT.

To investigate the amplitudes in Eqs.~\eqref{eq:ampsRR}--\eqref{eq:ampsLR}, we
utilize the crossing symmetry that shows that Eqs.~\eqref{eq:ampsRR} and
\eqref{eq:ampsLL}, as well as Eqs.~\eqref{eq:ampsRL} and \eqref{eq:ampsLR}, are
equal. Furthermore, Eqs.~\eqref{eq:ampsRR} and \eqref{eq:ampsRL} are related by
an $s\leftrightarrow u$ crossing,
\begin{flalign}
	&M(f_{1,R}f_{2,R} \to f_{1,R}f_{2,R})= C_{s}s^2 + C_{t} t^2 + C_{u} u^2,
	\\
	&M(f_{1,R}f_{2,L} \to f_{1,R}f_{2,L})= C_{s}u^2 + C_{t} t^2 + C_{u} s^2.
\end{flalign}
where $C_{s,t,u}$ denote generic coefficients. Since in the $t\to 0$ limit,
$u^2=s^2$, all four amplitudes lead to the same positivity
bound $C_{s}+C_{u}>0$. THus, it is sufficient to focus on only one of the
four cases.

Hence, we consider the $M(f_{1,R}f_{2,R} \to f_{1,R}f_{2,R})$ amplitude.
In the current problem, the right-handed $f_R$ fields consist of mixtures of
$e_R$, $\bar e_L$, and $\bar\nu_L$,
\begin{flalign}
	f_{n,R}=a_n e_R+b_n \bar e_L+c_n \bar\nu_L \quad \text{for}\quad n=1,2 ,
	\label{eq:mixing}
\end{flalign}
where $a_n,b_n,c_n$ are arbitrary complex numbers.
Using the results presented in Appendix~\ref{sec:app2}, the second-order
derivative of the amplitude w.r.t.~$s$ is determined to be
\be\bsp
  & \frac{1}{2}\frac{\ud^2}{\ud s^2}M(f_{1,R}f_{2,R} \to f_{1,R}f_{2,R}) =\\
  &\qquad Af_A+Bf_B+Cf_C+Df_D+Ef_E  ,
\esp\label{eq:positivityamp}\ee
where the quantities $f_A$, $f_B$, $f_C$, $f_D$, and $f_E$ are given by
\be\bsp
  & f_A\!=\!-4C_1,\ \ f_B\!=\!2C_2+C_3,\ \ f_C\!=\!-4(C_4\!+\!C_5),\\
  & f_D\!=\!-8C_5,\ \ f_E\!=\!C_3,\\
\esp\label{eq:coefs}\ee
and the parameters $A$, $B$, $C$, $D$, and $E$ are
\be\bsp
  & A \!=\! |a_1a_2^\dagger|^2, \ \
    B \!=\! 2 \Re (a_1a_2^\dagger)(\vec x^\dagger\!\cdot\!\vec y),\ \
    C \!=\! \left| \vec x \cdot \vec y^\dagger \right|^2, \\
  & D \!=\! \left| \vec x \times \vec y \right|^2, \ \
    E \!=\! |a_1|^2\left|\vec y\right|^2 + |a_2|^2\left|\vec x\right|^2\ .
\esp\ee
In these expressions, the vectors $\vec x$ and $\vec y$ and their 
various dot and cross products are defined as
\be\bsp
 & \vec x\equiv(b_1,c_1), \ \ \vec y\equiv(b_2,c_2),\\
 & \vec u\cdot \vec v \equiv u_1v_1+u_2v_2,\ \ \vec u\times \vec v \equiv u_1v_2-u_2v_1\ .
\esp\ee

The positivity bounds arise from the requirement that
Eq.~\eqref{eq:positivityamp} be positive for all possible values of
the parameters $A$, $B$, $C$, $D$, and $E$. It can be observed that
\begin{flalign}
	A>0,\quad C>0,\quad D>0  . \label{eq:AB1}
\end{flalign}
In addition,
\be
  \hspace*{-.2cm}
  |B| \!=\! 2 \left|\Re (a_1a_2^\dagger)\left( \vec x^\dagger\!\cdot\!\vec y \right)\right|
  \!<\! 2\left|a_1a_2^\dagger\right| \left|\vec x^\dagger\!\cdot\!\vec y \right| \!=\! 2\sqrt{AC},
\label{eq:AB2}\ee
and
\be
  E\!=\!|a_1|^2\left|\vec y\right|^2 \!+\! |a_2|^2\left|\vec x\right|^2
  \!>\! 2|a_1|\left|\vec y\right| |a_2|\left|\vec x\right| \!=\! 2\sqrt{A(C+D)},
\label{eq:AB3}\ee
where we have used $C+D=\left|\vec x\right|^2\left|\vec y\right|^2$.

We must then show that, for each set of parameters $A$, $B$, $C$, $D$, and $E$
that satisfy the inequalities of Eqs.~\eqref{eq:AB1}--\eqref{eq:AB3}, there
exists a corresponding set of coefficients $a_n$, $b_n$, and $c_n$. This is
achieved by introducing $\phi=\cos^{-1}\frac{B}{2\sqrt{AC}}$ and $r>1$ such that
$r^2+r^{-2}=E/\sqrt{A(C+D}$. Then, at least the following values for $a_n$,
$b_n$, and $c_n$ can be determined:
\be\bsp
&a_1=rA^{1/4}e^{i\phi},\\
&a_2=r^{-1}A^{1/4},\\
&(b_1,c_1)=\sqrt[4]{C+D}(1,0),\\
&(b_2,c_2)=\sqrt[4]{C+D}\left( \frac{C^{1/2}}{(C+D)^{1/2}},\frac{D^{1/2}}{(C+D)^{1/2}}\right) .
\esp\ee

Therefore, we conclude that the positivity bounds require that
\begin{flalign}
	f\equiv Af_A+Bf_B+Cf_C+Df_D+Ef_E >0 ,
\end{flalign}
for any $A,B,C,D,E$ that are real and that satisfy
\be\bsp
&\hspace*{1.8cm}A>0,\quad C>0, \quad D>0,\\
& -2\sqrt{AC}<B<2\sqrt{AC}, \quad 2\sqrt{A(C+D)}<E.
\esp\label{eq:range}\ee
Equivalently, this implies that the function $f$ has a minimum in the domain
defined by Eq.~\eqref{eq:range}, and that this minimum is positive. The existence
of a minimum implies that
\begin{flalign}
	f_D>0,\quad f_E>0 \label{eq:neqa} .
\end{flalign}
Moreover, as $|B|$ has an upper bound and $E$ has a lower bound, the minimum of
$f$ obtained by varying $B$ and $E$ is realized by
$B\to -\mbox{sign}(f_B)2\sqrt{AC}$ and $E\to 2\sqrt{A(C+D)}$. In other words,
\begin{flalign}
	&f>f_{BE}
	\equiv Af_A+Cf_C-2|f_B|\sqrt{AC}+Df_D 
	\nonumber\\
	&~~~~~~~~~~~~~~~~~~~~~~~~~~+2\sqrt{A(C+D)}f_E  .
\end{flalign}
Since $f_D>0$, we could further decrease this function by taking $D\to0$,
\begin{flalign}
	f>f_{BED}=Af_A + 2(f_E-|f_B|)\sqrt{AC}+Cf_C .
\end{flalign}
Finally, the minimal value of the above function has to be positive for any $A,C>0$, so that
\begin{flalign}
	&f_A>0,~~~~ f_C>0, \label{eq:neqb}\\
	&f_E>|f_B|\quad\mbox{or}\quad f_Af_C>(|f_B|-f_E)^2.
\end{flalign}
The last of which is equivalent to
\begin{flalign}
	\sqrt{f_Af_C}>f_B-f_E\quad\mbox{and}\quad
	\sqrt{f_Af_C}>-f_B-f_E  . \label{eq:neqc}
\end{flalign}

Combining the inequalities of Eqs.~\eqref{eq:neqa}, \eqref{eq:neqb}, and
\eqref{eq:neqc} and plugging in the actual Wilson coefficients from
Eq.~\eqref{eq:coefs}, we obtain the positivity bounds of
Eqs.~\eqref{eq:bound0}--\eqref{eq:bound2}.

\clearpage

\section{Amplitudes used for deriving positivity bounds}
\label{sec:app2}
{
\small
\vspace{-10pt}
\begin{flalign}
	\begin{aligned}
&M(e_Re_R\to e_Re_R)=-4 s^2 C_1 \\
 &M(e_Re_L\to e_Re_L)=C_3 (s+t)^2+2 t C_2 (s+t) \\
 &M(e_Re_L\to e_Le_R)=C_3 t^2+2 (s+t) C_2 t \\
 &M(e_R\nu _L\to e_R\nu _L)=C_3 (s+t)^2+2 t C_2 (s+t) \\
 &M(e_R\nu _L\to \nu _Le_R)=C_3 t^2+2 (s+t) C_2 t \\
 &M(e_R\bar{e}_R\to e_R\bar{e}_R)=-4 (s+t)^2 C_1 \\
 &M(e_R\bar{e}_R\to e_L\bar{e}_L)=t^2 C_3-2 s t C_2 \\
 &M(e_R\bar{e}_R\to \nu _L\bar{\nu }_L)=t^2 C_3-2 s t C_2 \\
 &M(e_R\bar{e}_R\to \bar{e}_Re_R)=-4 t^2 C_1 \\
 &M(e_R\bar{e}_R\to \bar{e}_Le_L)=C_3 (s+t)^2+2 s C_2 (s+t) \\
 &M(e_R\bar{e}_R\to \bar{\nu }_L\nu _L)=C_3 (s+t)^2+2 s C_2 (s+t) \\
 &M(e_R\bar{e}_L\to e_R\bar{e}_L)=s^2 C_3-2 s t C_2 \\
 &M(e_R\bar{e}_L\to \bar{e}_Le_R)=C_3 s^2+2 (s+t) C_2 s \\
 &M(e_R\bar{\nu }_L\to e_R\bar{\nu }_L)=s^2 C_3-2 s t C_2 \\
 &M(e_R\bar{\nu }_L\to \bar{\nu }_Le_R)=C_3 s^2+2 (s+t) C_2 s \\
 &M(e_Le_R\to e_Re_L)=C_3 t^2+2 (s+t) C_2 t \\
 &M(e_Le_R\to e_Le_R)=C_3 (s+t)^2+2 t C_2 (s+t) \\
 &M(e_Le_L\to e_Le_L)=-4 C_4 s^2-4 C_5 s^2 \\
 &M(e_L\nu _L\to e_L\nu _L)=4 s t C_4-4 s (2 s+3 t) C_5 \\
 &M(e_L\nu _L\to \nu _Le_L)=4 s (s+3 t) C_5-4 s (s+t) C_4 \\
 &M(e_L\bar{e}_R\to e_L\bar{e}_R)=s^2 C_3-2 s t C_2 \\
 &M(e_L\bar{e}_R\to \bar{e}_Re_L)=C_3 s^2+2 (s+t) C_2 s \\
 &M(e_L\bar{e}_L\to e_R\bar{e}_R)=t^2 C_3-2 s t C_2 \\
 &M(e_L\bar{e}_L\to e_L\bar{e}_L)=-4 C_4 (s+t)^2-4 C_5 (s+t)^2 \\
 &M(e_L\bar{e}_L\to \nu _L\bar{\nu }_L)=4 (s-2 t) (s+t) C_5-4 s (s+t) C_4 \\
 &M(e_L\bar{e}_L\to \bar{e}_Re_R)=C_3 (s+t)^2+2 s C_2 (s+t) \\
 &M(e_L\bar{e}_L\to \bar{e}_Le_L)=-4 C_4 t^2-4 C_5 t^2 \\
 &M(e_L\bar{e}_L\to \bar{\nu }_L\nu _L)=4 s t C_4-4 t (3 s+2 t) C_5 \\
 &M(e_L\bar{\nu }_L\to e_L\bar{\nu }_L)=-4 t (s+t) C_4-4 (2 s-t) (s+t) C_5 \\
 &M(e_L\bar{\nu }_L\to \bar{\nu }_Le_L)=4 t (3 s+t) C_5-4 t (s+t) C_4 \\
 &M(\nu _Le_R\to e_R\nu _L)=C_3 t^2+2 (s+t) C_2 t \\
 &M(\nu _Le_R\to \nu _Le_R)=C_3 (s+t)^2+2 t C_2 (s+t) \\
 &M(\nu _Le_L\to e_L\nu _L)=4 s (s+3 t) C_5-4 s (s+t) C_4 \\
 &M(\nu _Le_L\to \nu _Le_L)=4 s t C_4-4 s (2 s+3 t) C_5 \\
 &M(\nu _L\nu _L\to \nu _L\nu _L)=-4 C_4 s^2-4 C_5 s^2 \\
 &M(\nu _L\bar{e}_R\to \nu _L\bar{e}_R)=s^2 C_3-2 s t C_2 \\
 &M(\nu _L\bar{e}_R\to \bar{e}_R\nu _L)=C_3 s^2+2 (s+t) C_2 s \\
 &M(\nu _L\bar{e}_L\to \nu _L\bar{e}_L)=-4 t (s+t) C_4-4 (2 s-t) (s+t) C_5 \\
 &M(\nu _L\bar{e}_L\to \bar{e}_L\nu _L)=4 t (3 s+t) C_5-4 t (s+t) C_4 \\
 &M(\nu _L\bar{\nu }_L\to e_R\bar{e}_R)=t^2 C_3-2 s t C_2 \\
 &M(\nu _L\bar{\nu }_L\to e_L\bar{e}_L)=4 (s-2 t) (s+t) C_5-4 s (s+t) C_4 \\
 &M(\nu _L\bar{\nu }_L\to \nu _L\bar{\nu }_L)=-4 C_4 (s+t)^2-4 C_5 (s+t)^2 \\
 &M(\nu _L\bar{\nu }_L\to \bar{e}_Re_R)=C_3 (s+t)^2+2 s C_2 (s+t) \\
 &M(\nu _L\bar{\nu }_L\to \bar{e}_Le_L)=4 s t C_4-4 t (3 s+2 t) C_5 \\
 &M(\nu _L\bar{\nu }_L\to \bar{\nu }_L\nu _L)=-4 C_4 t^2-4 C_5 t^2 \\
 \phantom{xxxx}
\end{aligned}
\begin{aligned}
 &M(\bar{e}_Re_R\to e_R\bar{e}_R)=-4 t^2 C_1 \\
 &M(\bar{e}_Re_R\to e_L\bar{e}_L)=C_3 (s+t)^2+2 s C_2 (s+t) \\
 &M(\bar{e}_Re_R\to \nu _L\bar{\nu }_L)=C_3 (s+t)^2+2 s C_2 (s+t) \\
 &M(\bar{e}_Re_R\to \bar{e}_Re_R)=-4 (s+t)^2 C_1 \\
 &M(\bar{e}_Re_R\to \bar{e}_Le_L)=t^2 C_3-2 s t C_2 \\
 &M(\bar{e}_Re_R\to \bar{\nu }_L\nu _L)=t^2 C_3-2 s t C_2 \\
 &M(\bar{e}_Re_L\to e_L\bar{e}_R)=C_3 s^2+2 (s+t) C_2 s \\
 &M(\bar{e}_Re_L\to \bar{e}_Re_L)=s^2 C_3-2 s t C_2 \\
 &M(\bar{e}_R\nu _L\to \nu _L\bar{e}_R)=C_3 s^2+2 (s+t) C_2 s \\
 &M(\bar{e}_R\nu _L\to \bar{e}_R\nu _L)=s^2 C_3-2 s t C_2 \\
 &M(\bar{e}_R\bar{e}_R\to \bar{e}_R\bar{e}_R)=-4 s^2 C_1 \\
 &M(\bar{e}_R\bar{e}_L\to \bar{e}_R\bar{e}_L)=C_3 (s+t)^2+2 t C_2 (s+t) \\
 &M(\bar{e}_R\bar{e}_L\to \bar{e}_L\bar{e}_R)=C_3 t^2+2 (s+t) C_2 t \\
 &M(\bar{e}_R\bar{\nu }_L\to \bar{e}_R\bar{\nu }_L)=C_3 (s+t)^2+2 t C_2 (s+t)
   \\
 &M(\bar{e}_R\bar{\nu }_L\to \bar{\nu }_L\bar{e}_R)=C_3 t^2+2 (s+t) C_2 t \\
 &M(\bar{e}_Le_R\to e_R\bar{e}_L)=C_3 s^2+2 (s+t) C_2 s \\
 &M(\bar{e}_Le_R\to \bar{e}_Le_R)=s^2 C_3-2 s t C_2 \\
 &M(\bar{e}_Le_L\to e_R\bar{e}_R)=C_3 (s+t)^2+2 s C_2 (s+t) \\
 &M(\bar{e}_Le_L\to e_L\bar{e}_L)=-4 C_4 t^2-4 C_5 t^2 \\
 &M(\bar{e}_Le_L\to \nu _L\bar{\nu }_L)=4 s t C_4-4 t (3 s+2 t) C_5 \\
 &M(\bar{e}_Le_L\to \bar{e}_Re_R)=t^2 C_3-2 s t C_2 \\
 &M(\bar{e}_Le_L\to \bar{e}_Le_L)=-4 C_4 (s+t)^2-4 C_5 (s+t)^2 \\
 &M(\bar{e}_Le_L\to \bar{\nu }_L\nu _L)=4 (s-2 t) (s+t) C_5-4 s (s+t) C_4 \\
 &M(\bar{e}_L\nu _L\to \nu _L\bar{e}_L)=4 t (3 s+t) C_5-4 t (s+t) C_4 \\
 &M(\bar{e}_L\nu _L\to \bar{e}_L\nu _L)=-4 t (s+t) C_4-4 (2 s-t) (s+t) C_5 \\
 &M(\bar{e}_L\bar{e}_R\to \bar{e}_R\bar{e}_L)=C_3 t^2+2 (s+t) C_2 t \\
 &M(\bar{e}_L\bar{e}_R\to \bar{e}_L\bar{e}_R)=C_3 (s+t)^2+2 t C_2 (s+t) \\
 &M(\bar{e}_L\bar{e}_L\to \bar{e}_L\bar{e}_L)=-4 C_4 s^2-4 C_5 s^2 \\
 &M(\bar{e}_L\bar{\nu }_L\to \bar{e}_L\bar{\nu }_L)=4 s t C_4-4 s (2 s+3 t)
   C_5 \\
 &M(\bar{e}_L\bar{\nu }_L\to \bar{\nu }_L\bar{e}_L)=4 s (s+3 t) C_5-4 s (s+t)
   C_4 \\
 &M(\bar{\nu }_Le_R\to e_R\bar{\nu }_L)=C_3 s^2+2 (s+t) C_2 s \\
 &M(\bar{\nu }_Le_R\to \bar{\nu }_Le_R)=s^2 C_3-2 s t C_2 \\
 &M(\bar{\nu }_Le_L\to e_L\bar{\nu }_L)=4 t (3 s+t) C_5-4 t (s+t) C_4 \\
 &M(\bar{\nu }_Le_L\to \bar{\nu }_Le_L)=-4 t (s+t) C_4-4 (2 s-t) (s+t) C_5 \\
 &M(\bar{\nu }_L\nu _L\to e_R\bar{e}_R)=C_3 (s+t)^2+2 s C_2 (s+t) \\
 &M(\bar{\nu }_L\nu _L\to e_L\bar{e}_L)=4 s t C_4-4 t (3 s+2 t) C_5 \\
 &M(\bar{\nu }_L\nu _L\to \nu _L\bar{\nu }_L)=-4 C_4 t^2-4 C_5 t^2 \\
 &M(\bar{\nu }_L\nu _L\to \bar{e}_Re_R)=t^2 C_3-2 s t C_2 \\
 &M(\bar{\nu }_L\nu _L\to \bar{e}_Le_L)=4 (s-2 t) (s+t) C_5-4 s (s+t) C_4 \\
 &M(\bar{\nu }_L\nu _L\to \bar{\nu }_L\nu _L)=-4 C_4 (s+t)^2-4 C_5 (s+t)^2 \\
 &M(\bar{\nu }_L\bar{e}_R\to \bar{e}_R\bar{\nu }_L)=C_3 t^2+2 (s+t) C_2 t \\
 &M(\bar{\nu }_L\bar{e}_R\to \bar{\nu }_L\bar{e}_R)=C_3 (s+t)^2+2 t C_2 (s+t)
   \\
 &M(\bar{\nu }_L\bar{e}_L\to \bar{e}_L\bar{\nu }_L)=4 s (s+3 t) C_5-4 s (s+t)
   C_4 \\
 &M(\bar{\nu }_L\bar{e}_L\to \bar{\nu }_L\bar{e}_L)=4 s t C_4-4 s (2 s+3 t)
   C_5 \\
 &M(\bar{\nu }_L\bar{\nu }_L\to \bar{\nu }_L\bar{\nu }_L)=-4 C_4 s^2-4 C_5
   s^2 \\
   \phantom{xxxx}
\end{aligned}
\end{flalign}
}

\bibliography{refs}

\end{document}